\documentclass[pra,twocolumn,showpacs,preprintnumbers,superscriptaddress]
{revtex4-2}
\usepackage{times}
\usepackage{bm}
\usepackage{graphicx}
\usepackage{amsbsy}
\usepackage{amssymb}
\usepackage{amsmath}
\usepackage{amsfonts}
\usepackage{amsthm}
\usepackage{tikz}
\usepackage{caption}
\usepackage{refstyle}
\usepackage{braket}
\usepackage{subcaption}
\usepackage{comment}
\usepackage[colorlinks=true, citecolor=blue, linkcolor=blue, urlcolor=blue]{hyperref}
\usetikzlibrary{shapes, arrows, positioning, arrows, automata, quotes}
\begin{document}
\theoremstyle{plain}
\newtheorem{theorem}{Theorem}
\newtheorem{lemma}[theorem]{Lemma}
\newtheorem{corollary}[theorem]{Corollary}
\newtheorem{proposition}[theorem]{Proposition}
\newtheorem{conjecture}[theorem]{Conjecture}
\newtheorem{example}{Example}
\newtheorem{result}{Result}
\newcommand{\eqnref}[1]{Eq.~\eqref{#1}}
\theoremstyle{definition}
\newtheorem{definition}{Definition}
\theoremstyle{remark}
\newtheorem*{remark}{Remark}
\title{Combinatorial structures in quantum correlation: A new perspective}
\author{Rohit Kumar, Satyabrata Adhikari}
\email{rohitkumar@dtu.ac.in, satyabrata@dtu.ac.in} \affiliation{Delhi Technological
University, Delhi-110042, Delhi, India}
\begin{abstract}
Graph-theoretic structures play a central role in the description and analysis of quantum systems. In this work, we introduce a new class of quantum states, called $A_\alpha$-graph states, which are constructed from either unweighted or weighted graphs by taking the normalised convex combination of the degree matrix $D$ and the adjacency matrix $A_G$ of a graph $G$. The constructed states are different from the standard graph states arising from stabiliser formalism. Our approach is also different from the approach used by Braunstein et al. This class of states depend on a tunable mixing parameter $\alpha \in (0,1]$. We first establish the conditions under which the associated operator $\rho_\alpha^{A_G}$ is positive semidefinite and hence represents a valid quantum state. We then derive a positive partial transposition (PPT) condition for $A_{\alpha}$-graph states in terms of graph parameters. This PPT condition involves only the Frobenius norm of the adjacency matrix of the graph, the degrees of the vertices and the total number of vertices. For simple graphs, we obtain the range of the parameter $\alpha$ for which the $A_{\alpha}$-graph states represent a class of entangled states. We then develop a graph-theoretic formulation of a moments-based entanglement detection criterion, focusing on the recently proposed $p_3$-PPT criterion, which relies on the second and third moments of the partial transposition. Since the estimation of these moments is experimentally accessible via randomised measurements, swap operations, and machine-learning-based protocols, our approach provides a physically relevant framework for detecting entanglement in structured quantum states derived from graphs. This work bridges graph theory and moments-based entanglement detection, offering a new perspective on the role of combinatorial structures in quantum correlations.
\end{abstract}
 \maketitle
\section{Introduction}
Entanglement is the most fundamental resource in quantum information theory. It has applications ranging from quantum computation and communication to quantum metrology. Detecting and characterising quantum entanglement in composite quantum systems remains a challenging problem. The positive partial transpose criterion (PPT criterion \cite{Peres1996,Horodecki1996}) is among the most widely used theoretical tools for entanglement detection. It provides a necessary condition for separability in bipartite systems. Although the partial transposition map is positive but it is not completely positive, and therefore it can not be implemented directly as a physical operation \cite{Peres1996}. However, it has been shown that moments of the partially transposed density matrix can be experimentally estimated \cite{elben2020}. Several methods have been developed for such measurements. A few examples of such methods are randomised single-qubit measurements within the classical shadow formalism \cite{Huang2020}, multi-copy protocols involving cyclic shift or swap operators \cite{Ekert2002, Elben2018}, and, more recently, machine-learning-based schemes capable of estimating moments of arbitrary order \cite{Gray2018,Lin2023}. Motivated by these developments, Elben et al. \cite{elben2020} proposed a moment-based entanglement detection criterion (known as $p_3$-PPT criterion), which detects bipartite entanglement using only the second and third order moments of the partially transposed density matrix. \\
Parallel to these advances, graph-theoretic methods have been extensively employed to analyse quantum correlations and structural properties of quantum states. Traditional graph states, constructed within the stabiliser formalism, have been extensively studied due to their relevance in measurement-based quantum computation and error correction \cite{Schlingemann2001,Dr2003,Raussendorf2003}. However, these states represent only a narrow subclass of graph-associated quantum states. 
Braunstein et al. \cite{Braunstein2006_1} have given a method to associate a simple graph with a mixed quantum state, by taking the normalised Laplacian corresponding to the adjacency matrix of the graph. The normalised Laplacian satisfy all the properties of a quantum state. So, they have introduced the concept of density matrices of graphs to study the graphical representation of quantum states and their properties. They have also studied the entanglement properties of the mixed density matrices obtained from the combinatorial Laplacian. A. Cabello et al. \cite{Cabello2014} utilised graph-based techniques to characterise correlations arising in non-contextual theories, quantum theory, and more general probabilistic frameworks. M. Ray et al. \cite{Ray2021} adopted a graph theoretic approach to identify quantum dimension witnesses and thereby determine the minimum Hilbert-space dimension required for implementing a given quantum task. The entanglement properties of grid states have been graphically characterised by J. Lockhart et al. \cite{Lockhart2018}. Furthermore, the separability problem for bipartite quantum states generated from graphs was investigated in \cite{Dutta2016}. The entanglement properties of a quantum state have been studied using the graph Laplacian in \cite{Kumar2022}. They apply the unital map to represent a quantum state $\rho$ as the sum of $L_{\rho}$ and $\rho$, where $L_{\rho}$ satisfy all the properties of the graph Laplacian. Hence, they represented the quantum state as a graph corresponding to $L_{\rho}$. Laplacian matrices corresponding to weighted digraphs with complex weights, represented as quantum states, were studied by Adhikari et al. \cite{Adhikari2017}.   

In this article, we introduce a new class of quantum states associated with a graph. We call it $A_{\alpha}$-graph states, which are fundamentally different from the standard graph states. Our approach is also different from the approach used by Braunstein et al. \cite{Braunstein2006_1}. Given a simple unweighted or weighted graph $G$ with adjacency matrix $A_G$ and degree matrix $D$, we define a family of density operators, denoted by $\rho_{\alpha}^{A_G}$, as a normalised convex combination of these matrices, parametrised by a mixing parameter $\alpha$. This construction provides a direct and flexible mapping from graph structure to quantum states, incorporating weighted connectivity and allowing continuous interpolation between degree-dominated and adjacency-dominated regimes. First, we characterise the range of the parameter $\alpha$ for which the operator $\rho_{\alpha}^{A_G}$ is positive semidefinite and hence represents a valid quantum state. We derive PPT criteria based on the Frobenius norm of the adjacency matrix of the graph using the result on the second-order moment of the partially transposed state, given in \cite{KUMAR2026131195}.  Thereafter, we obtain a condition on the mixing parameter $\alpha$ to determine the interval in which the given class represent a family of negative partial transpose entangled states or a family of positive partial transposed states. We then obtain the PPT criterion in terms of graph parameters. Our approach enables the detection of entanglement, in $A_{\alpha}$-graph states, directly in terms of graph-derived quantities.

This paper is organised into different sections and subsections. In section \ref{sec:Preliminary_Results}, we present some preliminary results useful in our study. Section \ref{Sec:Graph_Brief} give a brief review of Graph Theory. We present here the basic definitions and results on Graph Theory, which we use in later sections. In section \ref{Sec:PPT_Condition_1} we derive a condition on the parameter $\alpha$ that ensures the validity of the $A_{\alpha}$-graph states. Then we obtain a PPT condition based on the Frobenius norm of the adjacency matrix of a weighted graph $G$. The corollary of the result gives us an entanglement detection criterion for $A_{\alpha}$-graph states in terms of the graph properties. We then present some graphs and corresponding classes of valid quantum states. We also verify our results for these classes of quantum states obtained from the graphs. In section \ref{Sec:PPT_Cond_Simple_Graph}, we obtain the range of $\alpha$ for which the $A_{\alpha}$-graph states obtained from a simple unweighted graph represent a class of entangled states. This range of alpha depends only on the degrees of the vertices of the graph and the number of vertices in the graph. Later in this section, we provide some examples to support the results obtained in this section. In section \ref{Sec:Graph_Moments_Based_Entanglement} we formulate the $p_3$-PPT criterion for $A_{\alpha}$-graph states, in terms of the graph quantities. This result enables us to identify the entanglement in $A_{\alpha}$-graph states using the graph structure. We end this section with some examples supporting our results obtained in this section. For completeness, some proofs are put in the Appendix (section \ref{Sec:Appendix}).
\section{Preliminary Results}
\label{sec:Preliminary_Results}
In this section, we state a few results that would be needed in the subsequent sections.\\
\begin{result}
\cite{Horn1985}
\label{result:Weyl's_inequality}
    Let $M_{n}$ denotes the set of $n \times n$ Hermitian matrices. If $X, Y \in M_{n}$ and the eigenvalues $\lambda_{i}(X)$, $\lambda_{i}(Y)$ and $\lambda_{i}(X + Y )$ be arranged in an increasing order then for each $k = 1, 2, ...n$, the following inequality holds
\begin{eqnarray}
\lambda_{k}(X) + \lambda_{1}(Y ) \leq \lambda_{k}(X + Y ) \leq \lambda_{k}(X) + \lambda_{n}(Y )
\end{eqnarray}
where  $\lambda_{1}(Y )$, $\lambda_{n}(Y)$ denotes the minimum and maximum eigenvalues of Y respectively.
\end{result} 
\begin{result} \cite{KUMAR2026131195} 
\label{result:moment_ppt_cond}
If $p_2\left(\rho^{T_B}\right) \leq \frac{1}{d_1 d_2-1}$ where the quantum state $\rho$ is in $d_1 \otimes d_2$ dimensional system, then $\rho \in PPT$ (set of positive all partial transposed states).
\end{result}
\section{A Brief Review on Graph Theory}
\label{Sec:Graph_Brief}
Graph theory is a well-established branch of Mathematics in which we represent the objects by vertices (or nodes) and the connection between them by edges. In this section, we give the basics of Graph theory. We start with the definition of a graph.\\
A graph is a mathematical structure that is used to model pairwise relations or connections between objects. A graph $G$ consists of a set $V(G)$ of vertices (or nodes) and a set $E(G)$ of edges. Each edge connects a pair of vertices and it is represented by $e_{ij} \equiv (v_i, v_j)$ if it connects the vertices $v_i$ and $v_j$. If two vertices are connected by an edge, those vertices are called adjacent vertices. There are different variants of a graph, such as directed/undirected graph, weighted/unweighted graph, multigraph, pseudograph, simple graph and bipartite graph.\\
(i) \textbf{Directed/undirected graph:} Undirected graphs are those where edges have no direction. A graph with directed edges is called a digraph.\\ (ii) \textbf{Weighted/unweighted graph:} graphs with weighted edges are referred to as weighted graphs. Otherwise, the graphs are said to be unweighted. A weighted graph $G=(V,E, w)$ induces a weight function $w:E(G)\rightarrow \mathbb{R}$ that assign a real number (weight) to each edge. In particular, if the weight function map each edge to 1 i.e. if the weight of each edge is 1 then the graph may be considered as unweighted graph.\\
(iii) \textbf{Multigraph:} If there are multiple edges between two vertices in a graph then the corresponding graph is called a multigraph.\\
(iv) \textbf{Pseudograph:} In some cases, loops (same initial and final vertex of an edge) are also allowed. A graph with multiple edges and loops is called a Pseudograph.\\
(v) \textbf{Simple graph:} An undirected graph without loops or multiple edges is called a simple graph.\\
(vi) \textbf{Bipartite graph:} If the vertex set $V(G)$ of a graph $G$ can be partitioned into two disjoint subsets namely $V_1(G)$ and $V_2(G)$ such that, $E(G)=\{(u,v) \,|\, u\in V_1(G), v\in V_2(G)\}$ i.e. each edge has one vertex in $V_1(G)$ and the other vertex in $V_2(G)$, then the graph $G$ is called a bipartite graph.\\

In this article, we consider simple unweighted and weighted graphs.
We can represent a weighted graph with a square matrix of order $n\times n$, where $n$ is the number of vertices in the graph. This matrix is called the adjacency matrix of the graph $G$. So, adjacency matrix corresponding to a weighted graph $G$ is a square matrix $A_G$ whose $(i,j)^{th}$ entry $a_{ij}$ is defined as,
\[
a_{ij}=
\begin{cases}
w_{ij}, &\text{ if } (v_i,v_j)\in E(G)\\
0, &\text{ if } (v_i,v_j)\notin E(G)
\end{cases}
\]
where $w_{ij}$ is the weight of the edge $(v_i, v_j)$. If the graph is unweighted then we set $w_{ij}=1$, for all $(v_i, v_j)\in E(G)$. Henceforth, we refer to a simple unweighted graph simply as a graph and a simple weighted graph as a weighted graph.\\ 
The degree of a vertex $v$ in a graph $G$ is the number of vertices adjacent to $v$, i.e. number of edges incident on $v$. It is denoted by $deg(v)$ or by $d_v$ or by $d(v)$. In case of a weighted graph, the weighted degree (or strength) of a vertex $v$ is defined as follows,
\[
d_v=\sum_{u\in N(v)}w_{uv}
\]
where $N(v)$ is the set of all vertices adjacent to $v$, which is also called the neighbourhood of $v$.

Now our task is to define the partially transposed graph of a given graph $G$. To start with, let us consider the adjacency matrix $A_G$ of a weighted graph $G$. We first define the partial transpose of $G$ by treating the graph as a bipartite system. In this setting, the associated quantum state is regarded as a $d_1 \otimes d_2$ dimensional bipartite quantum state. To define the partial transpose graph denoted by $G^{T_B}$ of $G$, where the partial transposition operation is considered with respect to the second subsystem, we first define a partition of the vertex set of $G$. Let $V(G)$ be the vertex set of $G$ and let us choose $d_1$ disjoint subsets $V_1, V_2,\cdots, V_{d_1}$ of $V(G)$ in such a way that $V_1 \cup V_2\cup \cdots \cup V_{d_1}  =V(G)$, and $|V_1|=d_2=|V_2|=\ldots =|V_{d_1}|$. Therefore, $|V(G)|=d_1 d_2$. 
\begin{definition} 
\label{def:graph_partial_transpose}
The graph obtained by taking the partial transpose of the adjacency matrix $A_G$ of an unweighted or weighted graph $G$ with respect to the system $B$, by treating $A_G$ as a bipartite system of dimension $d_1 \otimes d_2$, is defined as the partial transpose of the graph $G$ with respect to the system $B$. We denote this graph by $G^{T_B}$. In other words, we can define the partial transposition operation on the graph $G$ by replacing all the edges $(v_{ik}, v_{jl})$, by   $(v_{il}, v_{jk})$ (along with the corresponding weights in weighted graphs) \cite{Dutta2016}. Here, $(V_1, V_2,\ldots, V_{d_1})$ is a partition of $V(G)$ and $v_{ik}$ represents the $k^{th}$ vertex of $V_i$, $i\in \{1, 2, \ldots , d_1\}$. 
\end{definition}
The above definition can also be extended to multi-partite systems.
\begin{example}
Let us consider the graph $G_1$ given in Fig. (\ref{Fig:graph_G_and_GTB}).
Let $V(G_1)=\{v_{00}, v_{01}, v_{10}, v_{11}\}$ be the vertex set of $G_1$. Let $V_0$ and $V_1$ be disjoint subsets of $V(G_1)$ such that $V_0=\{v_{00},v_{01}\}$ and $V_1=\{v_{10}, v_{11}\}$. Thus, $(V_0, V_1)$ forms a bipartition of $V(G_1)$. The adjacency matrix $A_{G_1}$ of $G_1$ is given by 
\[
A_{G_1}=
\begin{bmatrix}
0 & 1 & 1 & 1 \\
1 & 0 & 0 & 1 \\
1 & 0 & 0 & 0 \\
1 & 1 & 0 & 0 
\end{bmatrix}.
\]
The partial transpose $A_{G_1}^{T_B}$ of $A_{G_1}$ with respect to the second subsystem is given by,
\[
A_{G_1}^{T_B} =
\begin{bmatrix}
0 & 1 & 1 & 0 \\
1 & 0 & 1 & 1 \\
1 & 1 & 0 & 0 \\
0 & 1 & 0 & 0
\end{bmatrix}.
\]
Hence, by definition (\ref{def:graph_partial_transpose}), ${G_1}^{T_B}$ is given in Fig. (\ref{Fig:graph_G_and_GTB}). 
 \begin{figure}[h]
    \centering

    \begin{subfigure}{0.4\textwidth}
    \centering
    \begin{tikzpicture}[every node/.style={circle, draw}]
        \node (A1) at (0,1.2) {$v_{00}$};
        \node (A2) at (3,1.2) {$v_{01}$};
        \node (B1) at (0,-1.2) {$v_{10}$};
        \node (B2) at (3,-1.2) {$v_{11}$};
        \draw (A1) -- (B1);
        \draw (A2) -- (B2);
        \draw (A1) -- (A2);
        \draw (A1) -- (B2);
    \end{tikzpicture}
    \caption{\(G_1\)}
    \end{subfigure}
    \hfill
    \begin{subfigure}{0.4\textwidth}
    \centering
    \begin{tikzpicture}[every node/.style={circle, draw}]
         \node (A1) at (0,1.2) {$v_{00}$};
        \node (A2) at (3,1.2) {$v_{01}$};
        \node (B1) at (0,-1.2) {$v_{10}$};
        \node (B2) at (3,-1.2) {$v_{10}$};
        \draw (A1) -- (B1);
        \draw (A1) -- (A2);
        \draw (A2) -- (B1);
        \draw (A2) -- (B2);
    \end{tikzpicture}
    \caption{\({G_1}^{T_B}\)}
    \end{subfigure}
    \caption{(a) Graph \(G_1\) and (b) its partial transpose \({G_1}^{T_B}\).}
    \label{Fig:graph_G_and_GTB}
\end{figure}
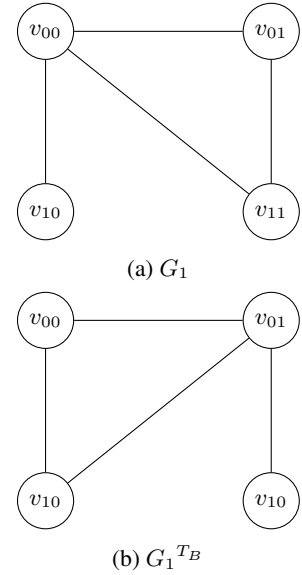
\\
Note that the edge set $E(G_1)$ of $G_1$ is given by,
\begin{align*}
E(G)=\{\, 
(v_{00},v_{01}),
(v_{00},v_{10}),
(v_{00},v_{11}),(v_{01},v_{11})\}
\end{align*}
Therefore, by the definition (\ref{def:graph_partial_transpose}), the edge set of $G^{T_B}$ will be,
\begin{align*}
E(G^{T_B})=\{ 
(v_{00},v_{01}),
(v_{00},v_{10}),
(v_{01},v_{10}),(v_{01},v_{11})\}
\end{align*}
\end{example}

\begin{example} Let us consider another graph $G_2$ with 9 vertices whose adjacency matrix is given by 

\[
A_{G_2} =
\begin{bmatrix}
0 & \tfrac{2}{5} & 0 & \tfrac{3}{4} & 0 & 0 & \tfrac{1}{2} & 0 & 0 \\[2pt]
\tfrac{2}{5} & 0 & \tfrac{7}{10} & 0 & \tfrac{1}{3} & 0 & 0 & 0 & \tfrac{4}{5} \\[2pt]
0 & \tfrac{7}{10} & 0 & \tfrac{1}{4} & \tfrac{2}{3} & 0 & 0 & \tfrac{3}{5} & 0 \\[2pt]
\tfrac{3}{4} & 0 & \tfrac{1}{4} & 0 & \tfrac{1}{2} & 0 & \tfrac{2}{5} & 0 & 0 \\[2pt]
0 & \tfrac{1}{3} & \tfrac{2}{3} & \tfrac{1}{2} & 0 & \tfrac{3}{4} & 0 & \tfrac{1}{5} & 0 \\[2pt]
0 & 0 & 0 & 0 & \tfrac{3}{4} & 0 & 0 & 0 & \tfrac{2}{3} \\[2pt]
\tfrac{1}{2} & 0 & 0 & \tfrac{2}{5} & 0 & 0 & 0 & \tfrac{4}{5} & 1 \\[2pt]
0 & 0 & \tfrac{3}{5} & 0 & \tfrac{1}{5} & 0 & \tfrac{4}{5} & 0 & \tfrac{7}{10} \\[2pt]
0 & \tfrac{4}{5} & 0 & 0 & 0 & \tfrac{2}{3} & 1 & \tfrac{7}{10} & 0
\end{bmatrix}.
\]
To define the partial transpose of $G_2$, we first treat $A_G$ as a bipartite system and then define a partition $P$ of $V(G_2)$. Let $(V_1, V_2, V_3)$ be a disjoint partition of $V(G)$, where $|V_i|=3, i=1,2,3$.
Let the vertex set is given by
\[
V(G_2)=\{v_{00}, v_{01}, v_{02}, v_{10}, v_{11}, v_{12}, v_{20}, v_{21}, v_{22}\}.
\]
where $v_{ik}$ represents the $k^{th}$ vertex of $V_i$.
and hence the edge set is
\[
\begin{aligned}
E(G_2)=\{\, &
(v_{00},v_{01},\tfrac{2}{5}),\,
(v_{00},v_{10},\tfrac{3}{4}),\,
(v_{00},v_{20},\tfrac{1}{2}),\\
& (v_{01},v_{02},\tfrac{7}{10}),\,
(v_{01},v_{11},\tfrac{1}{3}),\,
(v_{01},v_{22},\tfrac{4}{5}),\\
& (v_{02},v_{10},\tfrac{1}{4}),\,
(v_{02},v_{11},\tfrac{2}{3}),\,
(v_{02},v_{21},\tfrac{3}{5}),\\
& (v_{10},v_{11},\tfrac{1}{2}),\,
(v_{10},v_{20},\tfrac{2}{5}),(v_{11},v_{12},\tfrac{3}{4}),\\
&(v_{11},v_{21},\tfrac{1}{5}),
 (v_{12},v_{22},\tfrac{2}{3}),
 (v_{20},v_{21},\tfrac{4}{5}),\\
&(v_{20},v_{22},1),
(v_{21},v_{22},\tfrac{7}{10})
\,\}.
\end{aligned}
\]
Here, the third index $w_{ikjl}$ in $(v_{ik,v_{jl}, w_{ikjl}})$ represents the weight of the corresponding edge. Now, we apply the definition (\ref{def:graph_partial_transpose}) to obtain the edge set $E(G_2^{T_B})$ of $G_2^{T_B}$ as 
\[
\begin{aligned}
E(G_2^{T_B})=\{\, &
(v_{00},v_{01},\tfrac{2}{5}),\,
(v_{00},v_{10},\tfrac{3}{4}),\,
(v_{00},v_{20},\tfrac{1}{2}),\\
& (v_{01},v_{02},\tfrac{7}{10}),\,
(v_{01},v_{11},\tfrac{1}{3}),\,
(v_{02},v_{21},\tfrac{4}{5}),\\
& (v_{00},v_{12},\tfrac{1}{4}),\,
(v_{01},v_{12},\tfrac{2}{3}),\,
(v_{01},v_{22},\tfrac{3}{5}),\\
& (v_{10},v_{11},\tfrac{1}{2}),\,
(v_{10},v_{20},\tfrac{2}{5}),(v_{11},v_{12},\tfrac{3}{4}),\\
&(v_{11},v_{21},\tfrac{1}{5}),
 (v_{12},v_{22},\tfrac{2}{3}),
 (v_{20},v_{21},\tfrac{4}{5}),\\
&(v_{20},v_{22},1),
(v_{21},v_{22},\tfrac{7}{10})
\,\}.
\end{aligned}
\]
The corresponding adjacency matrix i.e. the adjacency matrix of $G_2^{T_B}$ with respect to the partition $P$ is given by
\[
A_{G_2^{T_B}}=
\begin{bmatrix}
0 & \tfrac{2}{5} & 0 & \tfrac{3}{4} & 0 & \tfrac{1}{4} & \tfrac{1}{2} & 0 & 0 \\[2pt]
\tfrac{2}{5} & 0 & \tfrac{7}{10} & 0 & \tfrac{1}{3} & \tfrac{2}{3} & 0 & 0 & \tfrac{3}{5} \\[2pt]
0 & \tfrac{7}{10} & 0 & 0 & 0 & 0 & 0 & \tfrac{4}{5} & 0 \\[2pt]
\tfrac{3}{4} & 0 & 0 & 0 & \tfrac{1}{2} & 0 & \tfrac{2}{5} & 0 & 0 \\[2pt]
0 & \tfrac{1}{3} & 0 & \tfrac{1}{2} & 0 & \tfrac{3}{4} & 0 & \tfrac{1}{5} & 0 \\[2pt]
\tfrac{1}{4} & \tfrac{2}{3} & 0 & 0 & \tfrac{3}{4} & 0 & 0 & 0 & \tfrac{2}{3} \\[2pt]
\tfrac{1}{2} & 0 & 0 & \tfrac{2}{5} & 0 & 0 & 0 & \tfrac{4}{5} & 1 \\[2pt]
0 & 0 & \tfrac{4}{5} & 0 & \tfrac{1}{5} & 0 & \tfrac{4}{5} & 0 & \tfrac{7}{10} \\[2pt]
0 & \tfrac{3}{5} & 0 & 0 & 0 & \tfrac{2}{3} & 1 & \tfrac{7}{10} & 0
\end{bmatrix}
\]

\begin{figure}[ht]
\centering
\begin{subfigure}{0.5\textwidth}
\begin{tikzpicture}[scale=1.5,
  every node/.style={circle, draw},
    unchanged/.style={thick},
    changed/.style={thick, dashed, red},
weight/.style={midway, draw=none, shape=rectangle, fill=white, inner sep=1pt, font=\footnotesize}
]

\node (v00) at (0,2) {$v_{00}$};
\node (v01) at (2,2) {$v_{01}$};
\node (v02) at (4,2) {$v_{02}$};

\node (v10) at (0,0.5) {$v_{10}$};
\node (v11) at (2,0.5) {$v_{11}$};
\node (v12) at (4,0.5) {$v_{12}$};

\node (v20) at (0,-1) {$v_{20}$};
\node (v21) at (2,-1) {$v_{21}$};
\node (v22) at (4,-1) {$v_{22}$};

\draw[unchanged] (v00)--(v01) node[weight] {$\tfrac{2}{5}$};
\draw[unchanged] (v00)--(v10) node[weight] {$\tfrac{3}{4}$};
\draw[unchanged, bend right=30]
  (v00) to node [weight] {$\tfrac{1}{2}$} (v20);
\draw[unchanged] (v01)--(v02) node[weight] {$\tfrac{7}{10}$};
\draw[unchanged] (v10)--(v11) node[weight] {$\tfrac{1}{2}$};
\draw[unchanged] (v11)--(v12) node[weight,pos=0.3] {$\tfrac{3}{4}$};
\draw[unchanged] (v20)--(v21) node[weight] {$\tfrac{4}{5}$};
\draw[unchanged] (v21)--(v22) node[weight] {$\tfrac{7}{10}$};
\draw[unchanged] (v10)--(v20) node[weight] {$\tfrac{2}{5}$};
\draw[unchanged] (v01)--(v11) node[weight,pos=0.25] {$\tfrac{1}{3}$};
\draw[unchanged] (v11)--(v21) node[weight] {$\tfrac{1}{5}$};
\draw[unchanged] (v12)--(v22) node[weight] {$\tfrac{2}{3}$};
\draw[unchanged, bend right=30]
  (v20) to node [weight] {$1$} (v22);

\draw[changed] (v01)--(v22) node[weight,pos=0.7] {$\tfrac{4}{5}$};
\draw[changed] (v02)--(v10) node[weight,pos=0.7] {$\tfrac{1}{4}$};
\draw[changed] (v02)--(v11) node[weight,pos=0.3] {$\tfrac{2}{3}$};
\draw[changed] (v02)--(v21) node[weight, pos=.3] {$\tfrac{3}{5}$};
\end{tikzpicture}
\subcaption{$G_2$}
\end{subfigure}
\begin{subfigure}{0.5\textwidth}
\begin{tikzpicture}[scale=1.5,
  every node/.style={circle, draw},
    unchanged/.style={thick},
    changed/.style={thick, dashed, red},
weight/.style={midway, draw=none, shape=rectangle, fill=white, inner sep=1pt, font=\footnotesize}
]

\node (u00) at (0,2) {$v_{00}$};
\node (u01) at (2,2) {$v_{01}$};
\node (u02) at (4,2) {$v_{02}$};

\node (u10) at (0,0.5) {$v_{10}$};
\node (u11) at (2,0.5) {$v_{11}$};
\node (u12) at (4,0.5) {$v_{12}$};

\node (u20) at (0,-1) {$v_{20}$};
\node (u21) at (2,-1) {$v_{21}$};
\node (u22) at (4,-1) {$v_{22}$};

\draw[unchanged] (u00)--(u01) node[weight] {$\tfrac{2}{5}$};
\draw[unchanged] (u00)--(u10) node[weight] {$\tfrac{3}{4}$};
\draw[unchanged, bend right=30]
  (u00) to node [weight] {$\tfrac{1}{2}$} (u20);
\draw[unchanged] (u01)--(u02) node[weight] {$\tfrac{7}{10}$};
\draw[unchanged] (u10)--(u11) node[weight] {$\tfrac{1}{2}$};
\draw[unchanged] (u11)--(u12) node[weight,pos=0.25] {$\tfrac{3}{4}$};
\draw[unchanged] (u20)--(u21) node[weight] {$\tfrac{4}{5}$};
\draw[unchanged] (u21)--(u22) node[weight] {$\tfrac{7}{10}$};
\draw[unchanged] (u10)--(u20) node[weight] {$\tfrac{2}{5}$};
\draw[unchanged] (u01)--(u11) node[weight, pos=0.3] {$\tfrac{1}{3}$};
\draw[unchanged] (u11)--(u21) node[weight] {$\tfrac{1}{5}$};
\draw[unchanged] (u12)--(u22) node[weight] {$\tfrac{2}{3}$};
\draw[unchanged, bend right=30]
  (u20) to node [weight] {$1$} (u22);

\draw[changed] (u02)--(u21) node[weight,pos=0.7] {$\tfrac{4}{5}$};
\draw[changed] (u00)--(u12) node[weight,pos=0.3] {$\tfrac{1}{4}$};
\draw[changed] (u01)--(u12) node[weight,pos=0.3] {$\tfrac{2}{3}$};
\draw[changed] (u01)--(u22) node[weight, pos=.7] {$\tfrac{3}{5}$};

\end{tikzpicture}
\subcaption{$G_2^{T_B}$}
\end{subfigure}
\caption{Graph $G_2$ and its partial transpose graph}
\label{fig:G_2_PT}
\end{figure}
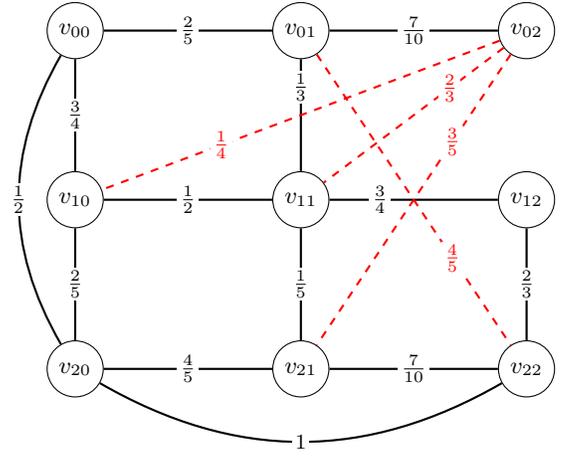
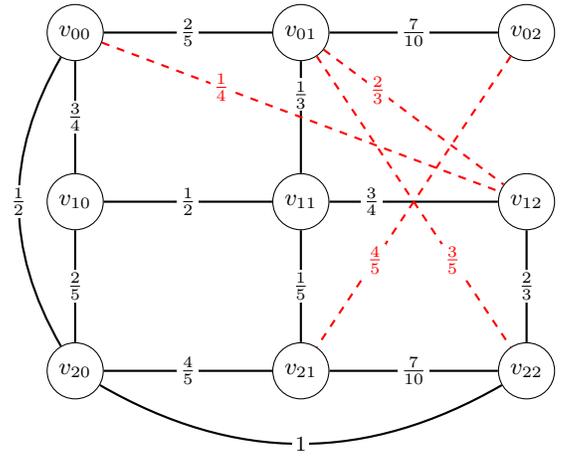

\end{example}
From the above examples, we can see that the partial transpose of a graph $G$ can be obtained by placing the vertices of $G$ in the grid formation (where rows represent a partition of the vertex set of $G$) and replacing the cross edges (edges connecting the vertices $v_{ik}$ and $v_{jl}$, $i\neq j$ and $k\neq l$) by their mirror images and keeping the edges between the same rows or columns unchanged.\\
We now derive the expressions for $\operatorname{Tr}\left(\left(A_{G_1}^{T_B}\right)^2\right)$ and $\operatorname{Tr}\left(\left(A_{G_1}^{T_B}\right)^3\right)$ in terms of graph theoretic parameters such as vertex degrees and edge weights.
\begin{lemma}
\label{lemma:Tr{(A^{T_B}})^2}
    Let $G$ be a weighted graph and $A_G$ be the weighted adjacency matrix of $G$, then,
    \begin{equation*}  \operatorname{Tr}\left(\left(A_G^{T_B}\right)^2\right)=2\sum_{(u,v)\in E(G)}w_{uv}^2=\|A_G\|_F^2
    \end{equation*}
    where $w_{uv}$ is the weight of the edge $(u,v)$ and $\|A_G\|_F$ represents the Frobenius norm of $A_G$. If $G$ is a simple unweighted graph, then,
    \begin{equation*}  \operatorname{Tr}\left(\left(A_G^{T_B}\right)^2\right)=2|E(G)|=\|A_G\|_F^2
    \end{equation*}
\end{lemma}
\begin{lemma}
\label{lemma:Tr_ATB_cube}
    Let $G$ be a weighted graph and $A_G$ be the weighted adjacency matrix of $G$, then,
    \begin{equation*}
        \operatorname{Tr}\left( (A^{T_B})^3 \right)=
6 \sum_{\triangle \in \mathcal{T}(G^{T_B})} \prod_{(u,v) \in \triangle} w_{uv}
    \end{equation*}
    where $\mathcal{T}(G^{T_B})$ is the set of all triangles (undirected 3-cycles) in the partial transpose graph $G^{T_B}$, and for each triangle $\triangle$ with edges $(u,v)$, the product is over the weights $w_{uv}$ of its three edges. If $G$ is a simple unweighted graph, then, 
    \[
    \operatorname{Tr}\left( (A^{T_B})^3 \right)=
6 \times \text{Number of triangles in $G^{T_B}$}
    \]
\end{lemma}
Proofs of the above results follow immediately from the well-known combinatorial interpretations of $\operatorname{Tr}((A_G)^2)$ and $\operatorname{Tr}((A_G)^3)$ \cite{west2000introduction,chung1997spectral}.
\begin{lemma}
    \label{lem:Tr(DA_TB_sq}
    Let $G$ be a weighted graph, $A$ be the weighted adjacency matrix of $G$, and $D$ be the weighted degree matrix (diagonal matrix with weighted degrees of each vertex on the diagonal). Let the vertex set $V(G)$ of $G$ be the disjoint union of $V_A$ and $V_B$, and a vertex can be represented by $v_{jk}$ if it is the $k^{th}$ vertex of the partition $V_j$, $j\in \{A, B\}$. Then,
    \begin{equation*}  \operatorname{Tr}\left(D\left(A^{T_B}\right)^2\right)=\sum_{(v_{ik},v_{jl})\in E(G)}\left(d_{v_{il}}+d_{v_{jk}}\right)(w_{v_{ik}v_{jl}})^2
    \end{equation*}
    where $d_{v_{ik}}$ represents the weighted degree of the $k^{th}$ vertex of the partition $V_i$.
\end{lemma}

The proof of the above lemma is given in Appendix in \hyperref[Sec:Appendix]{Section~\ref*{Sec:Appendix}}.

\section{\texorpdfstring{$A_{\alpha}-$graph State and PPT Condition}{Aalpha-graph State and PPT Condition}}
\label{Sec:PPT_Condition_1}
In this section, we will introduce a new family of quantum states that can be generated from the graph and then discuss its PPT criterion, which can be expressed in terms of the properties of a graph.
\subsection{$A_{\alpha}$- graph State} 
Let $G$ be an unweighted or weighted graph with edge weights $w_{ij} \in [0,1]$. If $A_{G}$ denote the adjacency matrix of the graph $G$ and $d_{G}$ denote the total weighted degree of $G$ defined as $d_G = \sum_{i=1}^{n} d_i$, where $d_i = \sum_{j=1}^{n} w_{ij}$ denote the weighted degree of the $i^\text{th}$ vertex then the linear combination of the degree matrix $D$ and the adjacency matrix $A_{G}$ may be denoted by $\rho_{\alpha}^{A_G}$ and can be expressed as
\begin{equation}
\rho_{\alpha}^{A_G} = \frac{1}{\alpha d_G} \left[ \alpha D + (1 - \alpha) A_{G} \right]
\label{rhodef}
\end{equation}
where $\alpha \in (0,1]$ denote the mixing parameter and can be chosen in such a way that $\rho_{\alpha}^{A_G}$ represent a positive semi-definite matrix. Therefore, our task is now reduces to finding the subinterval of $(0,1]$ in which $\rho_{\alpha}^{A_G}$ is positive semidefinite.\\
\begin{lemma}
\label{lemma:Postivity_cond}
$\rho_{\alpha}^{A_G}$ is positive semi-definite if the mixing parameter $\alpha$ is lying in the interval given by
\begin{eqnarray}
\frac{\lambda_{min}(A_{G})}{\lambda_{min}(A_{G})-\delta}\leq \alpha \leq 1
\end{eqnarray}
\end{lemma}
where $\lambda_{min}(A_{G})$ and $\delta$ respectively denote the minimum eigenvalue of the adjacency matrix and minimum degree of the graph $G$.\\ 
\textbf{Proof:} Let us recall $\rho_{\alpha}^{A_G}$ from (\ref{rhodef}). $\rho_{\alpha}^{A_G}$ is positive semi-definite if 
$\alpha D + (1 - \alpha) A_{G} \geq 0$ for some $\alpha$. Equivalently, we can express it as
\begin{equation}
D + \frac{1 - \alpha}{\alpha} A_{G} \geq 0
\label{ineq1}
\end{equation}
Since the matrix $D + \frac{1 - \alpha}{\alpha} A_{G}$ is hermitian so we apply Weyl's inequality given in result (\ref{result:Weyl's_inequality}) on it and thus we get
\begin{eqnarray}
\lambda_{\min}\Big(D + \frac{1-\alpha}{\alpha} A_{G} \Big) \ge \lambda_{\min}(D) + \lambda_{\min} \Big( \frac{1-\alpha}{\alpha} A_{G} \Big)
\label{weyl1}
\end{eqnarray}
We can now make an inference that $\rho_{\alpha}^{A_G}$ is positive semi-definite if 
\begin{eqnarray}
\lambda_{\min}(D) + \lambda_{\min} \left( \frac{1 - \alpha}{\alpha} A_{G} \right) \geq 0
\label{weyl2}
\end{eqnarray}
Since the degree matrix $D$ is a diagonal matrix so, we have $\lambda_{min}(D) = \delta.$\\
Therefore, (\ref{weyl2}) reduces to
\begin{eqnarray}
\alpha &\geq \frac{\lambda_{min}(A_{G})}{\lambda_{min}(A_{G})-\delta}
\end{eqnarray}
Thus, $\rho_{\alpha}^{A_G}$ is a positive definite matrix if $0<\frac{\lambda_{min}(A_{G})}{\lambda_{min}(A_{G})-\delta}\leq \alpha \leq 1$. Hence proved.

Therefore, $\rho_{\alpha}^{A_G}$ is a hermitian positive semidefinite matrix with unit trace if $\rho_{\alpha}^{A_G}$ represents a matrix of the form $\frac{1}{d_G}\left[D+\left((1-\alpha)/\alpha\right)A_{G}\right]$ with $\frac{\lambda_{min}(A_{G})}{\lambda_{min}(A_{G})-\delta}\leq \alpha \leq 1$ where $A_{G}$ is a weighted or simple graph adjacency matrix and $\delta$ is the minimum degree of graph $G$. Thus, $\rho_{\alpha}^{A_G}$ satisfies all the properties of a density matrix, and hence, a class of quantum states can be described by the density matrix $\rho_{\alpha}^{A_G}$. From now on, we will call this type of quantum state an $A_{\alpha}-$ graph state.\\
In particular, we note that if $\alpha=\frac{1}{2}$, then 
$\rho_{\frac{1}{2}}^{A_G}=\frac{1}{d_G}(D+A_{G})$, which is the normalised signless Laplacian matrix of $G$. An important observation we can make here is that a different vertex labelling would give rise to a different quantum state. Also, one may ask, what is the system dimension of the resultant quantum state? So, the answer to this question lies in the partitioning of the vertex set $V(G)$ of the graph $G$. 
\subsection{PPT Criterion in terms of the properties of $A_{\alpha}$-graph State}
We are now in a position to discuss the PPT criterion of $A_{\alpha}$-graph state. The interesting part of this section is that we have derived the PPT criterion completely in terms of the properties of a graph. In other words, we have shown that the PPT criterion of $A_{\alpha}$-graph state can be expressed in terms of a few properties of a graph $G$ such as the total degree $(d_{G})$ of the graph, degree $(d_{i})$ of the individual vertices of $G$ and the Frobenius norm of the adjacency matrix of the graph $G$.\\
To proceed towards our aim, let us start with the second-order moment of the partial transposition with respect to the subsystem $Y$ of the $m \otimes n$ dimensional bipartite system $XY$ described by the density matrix $\rho_{\alpha}^{A_G}$. The second order moment is denoted by $p_{2}(\rho_{\alpha}^{A_G})$ and it is given by
\begin{align}
p_2((\rho_{\alpha}^{A_G})^{T_{Y}}) 
&= \operatorname{Tr}\left[ \left( (\rho_{\alpha}^{A_G})^{T_Y} \right)^2 \right] \\
&= \frac{1}{(d_G)^2}\,
   \operatorname{Tr}\Bigg[
        D^2 
        + \left( \frac{1 - \alpha}{\alpha} \right)^2 (A_{G}^{T_Y})^2
        \\
&\quad +  \left (\frac{1 - \alpha}{\alpha}\right) (D A_{G}^{T_Y}+ A_{G}^{T_Y}D)
   \Bigg] \\
&= \frac{1}{(d_G)^2} \Bigg[
    \operatorname{Tr}(D^2)
    + \left( \frac{1 - \alpha}{\alpha} \right)^2 \operatorname{Tr}\!\big((A_{G}^{T_Y})^2\big)
       \Bigg]
\label{moment1}
\end{align}
In the last step of (\ref{moment1}), we have used the fact that $Tr(D A_{G}^{T_Y})=Tr(A_{G}^{T_Y} D)=0$. \\
Further, using the Lemma (\ref{lemma:Tr{(A^{T_B}})^2}), we get $\operatorname{Tr}\Big(\big(A_{G}^{T_B}\big)^2\Big)=\|A_{G}\|_F^2$, where $\|.\|_F$ represents the Frobenius norm.
Therefore, if $V(G)$ denote the set of all vertices of the graph $G$ and $d_{v}$ denote the degree of the vertex $v$ then (\ref{moment1}) reduces to
\begin{equation}
\label{eqn:p2}
 p_2((\rho_{\alpha}^{A_G})^{T_{Y}}) = \frac{1}{(d_G)^2} \left[
    \sum_{v\in V(G)} (d_v)^2
    + \left(\frac{1-\alpha}{\alpha}\right)^2\|A_{G}\|_F^2\right]
\end{equation}
If $m \otimes n$ dimensional quantum states described by the density matrix $\rho_{\alpha}^{A_G}$ and if the second order moment of the partial transposition with respect to the subsystem $Y$ of the density matrix $\rho_{\alpha}^{A_G}$ i.e.  $p_2((\rho_{\alpha}^{A_G})^{T_{Y}}) $ satisfies the following inequality 
\begin{eqnarray}
p_2((\rho_{\alpha}^{A_G})^{T_{Y}}) \leq \frac{1}{mn-1}
\label{ineq2}
\end{eqnarray}
then using Result-1, we can conclude that $\rho_{\alpha}^{A_G}$ represent a PPT state for some $\alpha$.\\
Using (\ref{ineq2}), the equation (\ref{eqn:p2}) can be expressed in the form of an inequality, which is given as 
\begin{eqnarray}
\frac{1}{(d_G)^2}
\left[
\sum_i (d_i)^2 +
\left(\frac{1 - \alpha}{\alpha}\right)^2
\|A_{G}\|_F^2
\right]
\leq \frac{1}{mn - 1} \nonumber\\
\Rightarrow \|A_{G}\|_F^2 \leq
\left(\frac{\alpha}{1 - \alpha}\right)^2
\left[
\frac{(d_G)^2}{mn - 1} -
\sum_i d_i^2
\right]
\end{eqnarray}
Thus, if 
$\|A_{G}\|_F^2 \leq
\left(\frac{\alpha}{1 - \alpha}\right)^2
\left[
\frac{(d_G)^2}{mn - 1} -
\sum_i d_i^2
\right]$
then $\rho_{\alpha}^{A_G}$ represent a PPT state.\\
The following theorem summarises the PPT criterion of $m \otimes n$ dimensional $A_{\alpha}$- graph state. 
\begin{theorem}
\label{thm: condition for PPT}
Let $G$ be a simple unweighted or a weighted graph and $A_{G}$ be its adjacency matrix. Further, assume that there exists some $\alpha$ for which $\rho_{\alpha}^{A_G}$ represent a family of $m \otimes n$ dimensional quantum states and they can be expressed in the form as
\[
\rho_{\alpha}^{A_G} = \frac{1}{\alpha d_G} \left[\alpha D + \left(1-\alpha\right) A_{G}\right],
\]
where $d_G = \sum_i d_i$, $d_i$ is the weighted degree of the $i^{\text{th}}$ vertex of $G$. Then the state $\rho_{\alpha}^{A_G}$ belong to  a family of PPT states if the following inequality  
\[
\|A_{G}\|_F^2 \leq \left(\frac{\alpha}{1 - \alpha}\right)^2
\left[
\frac{\left(d_G\right)^2}{mn - 1} -
\sum_i d_i^2
\right]
\]
holds.
\end{theorem}
\begin{corollary}
\label{cor:condition_for_PPT}
Let $\rho_{\alpha}^{A_G}$ be the $m \otimes n$ dimensional $A_\alpha-$ graph state corresponding to a weighted (or simple) graph $G$ as stated in Theorem (\ref {thm: condition for PPT}). If $\rho_{\alpha}^{A_G}$ represent a class of negative partial transpose entangled states then
\[
\|A_{G}\|_F^2 > \left(\frac{\alpha}{1 - \alpha}\right)^2
\left[
\frac{(d_G)^2}{mn - 1} -
\sum_i d_i^2
\right]
\]
\end{corollary}
\subsection{Examples-I}
In this section, we will verify the PPT criterion obtained in the earlier section with a few examples.
\begin{example}
Let us consider the graph $G_{1}$ given in Fig.1. 
\begin{figure}[h]
	\centering
	\begin{tikzpicture}
		\node[circle, draw] (1) at (0,0) {$v_{00}$};
		\node[circle, draw] (2) at (3,0) {$v_{01}$};
		\node[circle, draw] (3) at (0,-2) {$v_{10}$};
		\node[circle, draw] (4) at (3,-2) {$v_{11}$};
		
		\draw (1) -- (2) node[midway, fill=white] {$\tfrac{7}{50}$};
		\draw (1) -- (4) node[midway, fill=white, pos=0.3] {$\tfrac{1}{4}$};
		\draw (2) -- (4) node[midway, fill=white] {$\tfrac{1}{5}$};
		\draw (2) -- (3) node[midway, below, pos=0.7] {$\tfrac{9}{100}$};
	\end{tikzpicture}
	\caption{$G_1$}
	\label{fig:G1}
\end{figure}
The adjacency matrix $A_{G_1}$ corresponding to the graph $G_{1}$ can be represented as 
\begin{eqnarray}
A_{G_1} =
\begin{bmatrix}
	0 & \tfrac{7}{50} & 0 & \tfrac{1}{4} \\
	\tfrac{7}{50} & 0 & \tfrac{9}{100} & \tfrac{1}{5} \\
	0 & \tfrac{9}{100} & 0 & 0 \\
	\tfrac{1}{4} & \tfrac{1}{5} & 0 & 0
\end{bmatrix}
\end{eqnarray}
The minimum eigenvalue of $A_{G_{1}}$ is given by $\lambda_{min}(A_{G_{1}})= -0.2647$ and the minimum degree of the graph $G_{1}$ is $\delta=\frac{9}{100}$. Therefore, for $\alpha \in [0.75,1)$, $\rho_{\alpha}^{A_{G_{1}}}$ represent a family of states given as
\begin{eqnarray}
\rho_\alpha^{A_{G_1}} = \frac{1}{68}
\begin{bmatrix}
	\dfrac{39}{2} & 7 \beta & 0 & \dfrac{25}{2} \beta \\[8pt]
	7 \beta & \dfrac{43}{2} & \dfrac{9}{2} \beta & 10 \beta \\[8pt]
	0 & \dfrac{9}{2} \beta & \dfrac{9}{2} & 0 \\[8pt]
	\dfrac{25}{2} \beta & 10 \beta & 0 & \dfrac{45}{2}
\end{bmatrix}
\end{eqnarray}
where $\beta=\frac{1-\alpha}{\alpha}$.\\
By seeing the graph and its adjacency matrix, the quantities $\|A_{G_1}\|_F^2$ and $ \left(\frac{\alpha}{1 - \alpha}\right)^2 \left[\frac{(d_{G_1})^2}{mn - 1} -\sum_{v\in V(G_1)} d_v^2
\right]$ can be easily calculated as $\|A_{G_1}\|_F^2=0.26$ and $ \left(\frac{\alpha}{1 - \alpha}\right)^2 \left[\frac{(d_{G_1})^2}{mn - 1} -\sum_{v\in V(G_1)} d_v^2
\right]=\frac{517 \alpha ^2}{7500 (1-\alpha )^2}$. Therefore, it can be easily verified that the inequality $\|A_{G_1}\|_F^2 \leq  \left(\frac{\alpha}{1 - \alpha}\right)^2 \left[\frac{(d_G)^2}{mn - 1} -\sum_{v\in V(G_1)} d_v^2
\right]$ holds for $\alpha \in [0.75,1]$. Thus,
from Theorem (\ref{thm: condition for PPT}), the state $\rho_\alpha^{A_{G_1}}$ represent a family of $2 \otimes 2$ dimensional $PPT$ state for  $\alpha \in [0.75,1]$.
\end{example}
\begin{example}
Consider a graph $G_{2}$, which is given in Fig.2. 
\begin{figure}[h]
	\centering
	\begin{tikzpicture}[scale=1.5,
    node/.style={circle, draw},
    weight/.style={midway, font=\small,  fill=white, inner sep=1.4pt}
]
 \node[circle, draw] (1) at (0,0) {$v_{00}$};
    \node[circle, draw] (2) at (2,0) {$v_{01}$};
    \node[circle, draw] (3) at (4,0) {$v_{02}$};
    \node[circle, draw] (4) at (0,-2) {$v_{10}$};
     \node[circle, draw] (5) at (2,-2) {$v_{11}$};
    \node[circle, draw] (6) at (4,-2) {$v_{12}$};
    
		\draw (1) -- (2) node[midway,above] {$\tfrac{4}{5}$};
        \draw[bend left=30]
  (1) to node [weight] {$\tfrac{1}{20}$} (3);
		
		\draw (1) -- (5) node[midway, pos=0.2, left] {$\tfrac{73}{100}$};
		\draw (2) -- (4) node[midway, pos=0.75, left] {$\tfrac{1}{100}$};
		\draw (2) -- (6) node[midway, below] {$\tfrac{39}{100}$};	
		\draw (3) -- (4) node[midway, below] {$\tfrac{17}{50}$};
		\draw[bend right=30] (3) -- (6) node[midway, right] {$\tfrac{4}{25}$};	
		\draw (4) -- (5) node[midway, below] {$\tfrac{3}{4}$};
        \draw[bend right=30]
  (4) to node [weight] {$\tfrac{33}{50}$} (6);
		
	\end{tikzpicture}
	\caption{$G_2$}
	\label{fig:G2}
\end{figure}
The adjacency matrix $A_{G_2}$ of $G_{2}$ is given by
\begin{eqnarray}
A_{G_2} =
\begin{bmatrix}
	0 & \tfrac{4}{5} & \tfrac{1}{20} & 0 & \tfrac{73}{100} & 0 \\[6pt]
	\tfrac{4}{5} & 0 & 0 & \tfrac{1}{100} & 0 & \tfrac{39}{100} \\[6pt]
	\tfrac{1}{20} & 0 & 0 & \tfrac{17}{50} & 0 & \tfrac{4}{25} \\[6pt]
	0 & \tfrac{1}{100} & \tfrac{17}{50} & 0 & \tfrac{3}{4} & \tfrac{33}{50} \\[6pt]
	\tfrac{73}{100} & 0 & 0 & \tfrac{3}{4} & 0 & 0 \\[6pt]
	0 & \tfrac{39}{100} & \tfrac{4}{25} & \tfrac{33}{50} & 0 & 0
\end{bmatrix}
\end{eqnarray}
By calculating the minimum eigenvalue of $A_{G_{2}}$ and the minimum degree of $G_{2}$, we can calculate the range of the mixing parameter, which can be given as $\alpha \in [0.7,1]$. In this interval of $\alpha$, $\rho_{\alpha}^{A_{G_2}}$ represent a family of $2 \otimes 3$ dimensional quantum states, which is given by
\begin{eqnarray}
\rho_\alpha^{A_{G_2}} = \frac{1}{389}
\begin{bmatrix} \label{matrix:rhoG_2}
	79 & 40 \beta & \frac{5}{2} \beta & 0 & \frac{73}{2} \beta & 0 \\[6pt]
	40 \beta & 60 & 0 & \frac{1}{2} \beta & 0 & \frac{39}{2} \beta \\[6pt]
	\frac{5}{2} \beta & 0 & \frac{55}{2} & 17 \beta & 0 & 8 \beta \\[6pt]
	0 & \frac{1}{2} \beta & 17 \beta & 88 & \frac{75}{2} \beta & 33 \beta \\[6pt]
	\frac{73}{2} \beta & 0 & 0 & \frac{75}{2} \beta & 74 & 0 \\[6pt]
	0 & \frac{39}{2} \beta & 8 \beta & 33 \beta & 0 & \frac{121}{2}
\end{bmatrix}
\end{eqnarray}
where $\beta = \frac{1-\alpha}{\alpha}$.\\
For this class of quantum states described by the density matrix $\rho_\alpha^{A_{G_2}}$, we have
\begin{align}
\label{Exp:Example-2_Cond}
\begin{split}
\|A_{G_2}\|_F^2 - \left(\frac{\alpha}{1 - \alpha}\right)^2
\Biggl[\frac{(d_G)^2}{mn - 1} - \sum_{v \in V(G_2)} d_v^2 \Biggr] \\
= \frac{24669}{5000} - \frac{27867 \, \alpha^2}{25000 \, (1-\alpha)^2} \\
\end{split}
\end{align}
The right-hand side of equation (\ref{Exp:Example-2_Cond}) is less than zero for $\alpha \in \left[0.7,1\right)$. Therefore, by Theorem (\ref{thm: condition for PPT}) the class of quantum states $\rho_\alpha^{A_{G_2}}$  represents a class of PPT states for all values of $\alpha$ in the interval $\left[0.7,1\right)$.
Now, let us take one more example.    
\end{example}
\begin{example}
Let us consider a graph $G_{3}$ given in Fig.3. The corresponding adjacency matrix and the induced class of quantum states are given by 
\begin{eqnarray}
A_{G_3} =
\begin{bmatrix}
0 & \frac{13}{50} & 0 & \frac{4}{5} & 0 & \frac{29}{50} & \frac{41}{100} & \frac{12}{25} & 0 \\[6pt]
\frac{13}{50} & 0 & \frac{14}{25} & 0 & \frac{19}{25} & 0 & 0 & 0 & \frac{29}{50} \\[6pt]
0 & \frac{14}{25} & 0 & \frac{51}{100} & \frac{69}{100} & 0 & 0 & \frac{49}{100} & 0 \\[6pt]
\frac{4}{5} & 0 & \frac{51}{100} & 0 & \frac{19}{25} & 0 & \frac{12}{25} & 0 & 0 \\[6pt]
0 & \frac{19}{25} & \frac{69}{100} & \frac{19}{25} & 0 & 0 & \frac{9}{25} & \frac{11}{100} & 0 \\[6pt]
\frac{29}{50} & 0 & 0 & 0 & 0 & 0 & 0 & 0 & 0 \\[6pt]
\frac{41}{100} & 0 & 0 & \frac{12}{25} & \frac{9}{25} & 0 & 0 & \frac{64}{100} & 1 \\[6pt]
\frac{12}{25} & 0 & \frac{49}{100} & 0 & \frac{11}{100} & 0 & \frac{64}{100} & 0 & \frac{23}{25} \\[6pt]
0 & \frac{29}{50} & 0 & 0 & 0 & 0 & 1 & \frac{23}{25} & 0
\end{bmatrix}  
\label{ag3}
\end{eqnarray}
\begin{figure}[h]
\centering
\begin{tikzpicture}[scale=1.3,
    node/.style={circle, draw},
    weight/.style={midway, fill=white, font=\small, inner sep=1.4pt}
]

 \node[circle, draw] (00) at (0,0) {$v_{00}$};
    \node[circle, draw] (01) at (2,0) {$v_{01}$};
    \node[circle, draw] (02) at (4,0) {$v_{02}$};
    \node[circle, draw] (10) at (0,-2) {$v_{10}$};
     \node[circle, draw] (11) at (2,-2) {$v_{11}$};
    \node[circle, draw] (12) at (4,-2) {$v_{12}$};
    \node[circle, draw] (20) at (0,-4) {$v_{20}$};
     \node[circle, draw] (21) at (2,-4) {$v_{21}$};
    \node[circle, draw] (22) at (4,-4) {$v_{22}$};
\draw (00)--(01) node[weight] {$\tfrac{13}{50}$};
\draw (00)--(10) node[weight] {$\tfrac{4}{5}$};
\draw (00)--(12) node[weight, pos=0.3] {$\tfrac{29}{50}$};
\draw[bend right=30] (00) to node [weight] {$\tfrac{41}{100}$} (20);
\draw (00)--(21) node[weight, pos=0.3] {$\tfrac{12}{25}$};

\draw (01)--(02) node[weight] {$\tfrac{14}{25}$};
\draw (01)--(11) node[weight, pos=3.5] {$\tfrac{19}{25}$};
\draw (01)--(22) node[weight, pos=0.7] {$\tfrac{29}{50}$};

\draw (02)--(10) node[weight, pos=0.3] {$\tfrac{51}{100}$};
\draw (02)--(11) node[weight] {$\tfrac{69}{100}$};
\draw (02)--(21) node[weight, pos=0.7] {$\tfrac{49}{100}$};

\draw (10)--(11) node[weight, pos=0.7] {$\tfrac{19}{25}$};
\draw (10)--(20) node[weight] {$\tfrac{12}{25}$};

\draw (11)--(20) node[weight] {$\tfrac{9}{25}$};
\draw (11)--(21) node[weight] {$\tfrac{11}{100}$};

\draw (20)--(21) node[weight] {$\tfrac{64}{100}$};
\draw[bend right=30]
  (20) to node [weight] {$1$} (22);
  
\draw (21)--(22) node[weight] {$\tfrac{23}{25}$};

\end{tikzpicture}
\caption{$G_3$}
\label{fig:G3}
\end{figure}
\begin{equation}  
    \label{state:rho_A_G_3}
    \fontsize{8}{10}
\rho_\alpha^{A_{G_3}} =
\frac{1}{2078}
\begin{bmatrix}
253 & 26 \beta & 0 & 80 \beta & 0 & 58 \beta & 41 \beta & 48 \beta & 0 \\[6pt]
26 \beta & 216 & 56 \beta & 0 & 76 \beta & 0 & 0 & 0 & 58 \beta \\[6pt]
0 & 56 \beta & 225 & 51 \beta & 69 \beta & 0 & 0 & 49 \beta & 0 \\[6pt]
80 \beta & 0 & 51 \beta & 255 & 76 \beta & 0 & 48 \beta & 0 & 0 \\[6pt]
0 & 76 \beta & 69 \beta & 76 \beta & 268 & 0 & 36 \beta & 11 \beta & 0 \\[6pt]
58 \beta & 0 & 0 & 0 & 0 & 58 & 0 & 0 & 0 \\[6pt]
41 \beta & 0 & 0 & 48 \beta & 36 \beta & 0 & 289 & 32 \beta & 50 \beta \\[6pt]
48 \beta & 0 & 49 \beta & 0 & 11 \beta & 0 & 32 \beta & 264 & 46 \beta \\[6pt]
0 & 58 \beta & 0 & 0 & 0 & 0 & 50 \beta & 46 \beta & 250
\end{bmatrix}
\end{equation}
where $\beta=\frac{1-\alpha}{\alpha}$.\\
Let us now calculate the following expression by using the degree of the vertices of the graph and the adjacency matrix (\ref{ag3}), which is given by
\begin{align}
\label{Exp:Example-3_Cond}
\begin{split}
\|A_{G_3}\|_F^2 - \left(\frac{\alpha}{1 - \alpha}\right)^2
\Biggl[\frac{(d_G)^2}{mn - 1} - \sum_{v \in V(G_3)} d_v^2 \Biggr] \\
= \frac{68521}{5000}-\frac{45081 \alpha ^2}{20000 (1-\alpha )^2}\\
\end{split}
\end{align}
We can find that the right-hand side of equation (\ref{Exp:Example-3_Cond}) is less than zero for $\alpha \in \left[0.75,1\right)$. Therefore, by Theorem (\ref{thm: condition for PPT}), the class of quantum states given by $\rho_\alpha^{A_{G_3}}$  represents the PPT states for all values of $\alpha$ in the interval $\alpha \in \left[0.75,1\right)$.    
\end{example} 
\begin{example}
    Let us now consider the graph $G_4$ given in Fig. 4. The adjacency matrix corresponding to $G_4$ is given by,
\begin{figure}[h]
\centering
\begin{tikzpicture}[every node/.style={circle, draw}]
    \node (1) at (0,0) {$v_{00}$};
    \node (2) at (2,0) {$v_{01}$};
    \node (3) at (4,0) {$v_{02}$};
    \node (4) at (0,-2) {$v_{10}$};
    \node (5) at (2,-2) {$v_{11}$};
    \node (6) at (4,-2) {$v_{12}$};
    \draw (1) -- (2);
    \draw[bend left=30] (1) to (3);
    \draw (2) -- (5);
    \draw (2) -- (6);
    \draw (3) -- (4);
    \draw[bend right=30] (4) to (6);
    \draw (5) -- (6);
\end{tikzpicture}
 \caption{$G_4$}
\label{fig:G4}
\end{figure}
\[
A_{G_4}=
\left[
\begin{array}{cccccc}
 0 & 1 & 1 & 0 & 0 & 0 \\
 1 & 0 & 0 & 0 & 1 & 1 \\
 1 & 0 & 0 & 1 & 0 & 0 \\
 0 & 0 & 1 & 0 & 0 & 1 \\
 0 & 1 & 0 & 0 & 0 & 1 \\
 0 & 1 & 0 & 1 & 1 & 0 \\
\end{array}
\right]
\]
The corresponding $A_{\alpha}$-graph state is given by,
\[
\rho_{\alpha}^{A_{G_4}}=
\frac{1}{14}
\left[
\begin{array}{cccccc}
 2 & \beta & \beta & 0 & 0 & 0 \\
 \beta & 3 & 0 & 0 & \beta & \beta \\
 \beta & 0 & 2 & \beta & 0 & 0 \\
 0 & 0 & \beta & 2 & 0 & \beta \\
 0 & \beta & 0 & 0 & 2 & \beta \\
 0 & \beta & 0 & \beta & \beta & 3 \\
\end{array}
\right],\quad \beta=\frac{1-\alpha}{\alpha}
\]
Using Lemma (\ref{lemma:Postivity_cond}), we can show that $\rho_{\alpha}^{A_{G_4}}$ represent a valid class of quantum states for $\alpha \in [0.4676, 1]$. Using Peres-Horodecki PPT criterion, $\rho_{\alpha}^{A_{G_4}}$ is entangled for $\alpha \in [0.4676, 0.5247]$. Therefore, by Corollary (\ref{cor:condition_for_PPT})
\[
\|A_{G_2}\|_F^2 > \left(\frac{\alpha}{1 - \alpha}\right)^2
\left[
\frac{(d_{G_2})^2}{mn - 1} -
\sum_{v \in V(G_2)} d_v^2 \right]
\]
i.e.
\begin{equation}
\label{eqn:ent_cond}
\|A_{G_2}\|_F^2 -\left(\frac{\alpha}{1 - \alpha}\right)^2
\left[
\frac{(d_{G_2})^2}{mn - 1} -
\sum_{v\in V(G_2)}d_v^2 \right]>0
\end{equation}
Now, after the calculation, we obtain,
\begin{align*}
\|A_{G_2}\|_F^2 -\left(\frac{\alpha}{1 - \alpha}\right)^2
\left[
\frac{(d_{G_2})^2}{mn - 1} -
\sum_{v\in V(G_2)}d_v^2 \right] \\
=\frac{19 \alpha ^2}{2 (\alpha -1)^2}+14
\end{align*}
which is greater than 0 for $\alpha \in [0.4676, 0.5247]$. This implies that the equation (\ref{eqn:ent_cond}) is satisfied for $\alpha \in [0.4676, 0.5247]$ and hence the Corollary-(\ref{cor:condition_for_PPT}) is verified.
\end{example} 

\section{PPT Condition in terms of the characteristic of a simple unweighted graph}
\label{Sec:PPT_Cond_Simple_Graph}
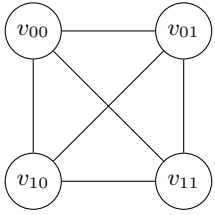
\begin{figure}[h]
\centering
\begin{tikzpicture}
    \node[circle, draw] (1) at (0,2) {$v_{00}$};
    \node[circle, draw] (2) at (2,2) {$v_{01}$};
    \node[circle, draw] (3) at (0,0) {$v_{10}$};
    \node[circle, draw] (4) at (2,0) {$v_{11}$};

    \draw (1) -- (2) node[midway, above] {};
     \draw (1) -- (3) node[midway, left] {};
      \draw (1) -- (4) node[pos=0.25, above] {};
    \draw (2) -- (3) node[pos=0.75, above] {};
     \draw (2) -- (4) node[midway, right] {};
    \draw (3) -- (4) node[midway, above] {};
\end{tikzpicture}
\caption{Complete graph $K_4$}
\label{fig:K4}
\end{figure}

In this section, we simplify the PPT conditions obtained in Theorem (\ref{thm: condition for PPT}) for simple graphs. The simplified PPT conditions put restrictions on the mixing parameter. We are then able to obtain the subinterval of the mixing parameter $\alpha$ for which the quantum state described by the density operator $\rho_{\alpha}^{A_{G}}$ corresponding to a simple graph $G$, represents a PPT state. 
\begin{theorem}
\label{thm:PPT_cond_alpha}
    Let $G$ be a graph, $A_G$ be its adjacency matrix, and $\rho_{\alpha}^{A_G}$ be the $m\otimes n$ dimensional $A_\alpha-$graph state corresponding to $G$. $ \rho_{\alpha}^{A_G}$ belongs to a class of PPT state if the following inequality holds
\begin{eqnarray}
\frac{1}{1+\sqrt{\frac{d_G}{mn-1}-\frac{1}{d_G}\sum_{v\in V(G)}d_v^2}}\leq \alpha <1  
\end{eqnarray}
\end{theorem}
\textbf{Proof:} Recalling Theorem (\ref{thm: condition for PPT}), we have $\rho_{\alpha}^{A_G} \in \text{PPT}$, if
\begin{align*}
\begin{split}
\|A\|_F^2 \leq \left(\frac{\alpha}{1 - \alpha}\right)^2
\left[
\frac{\left(d_G\right)^2}{mn - 1} -
\sum_{v\in V(G)}d_v^2
\right]
\end{split}
\end{align*}
For simple graphs, we have $\|A\|_F^2=d_G$.
Therefore, solving the above inequality in terms of the mixing parameter $\alpha$, we get
\begin{eqnarray}
\label{eqn:PPT_cond_simple}
      \frac{1}{1+\sqrt{\frac{d_G}{mn-1}-\frac{1}{d_G}\sum_{v\in V(G)}d_v^2}}\leq \alpha <1
\end{eqnarray}
Therefore, for a simple graph $G$, the quantum state $\rho_{\alpha}^{A_G}$ corresponding to $G$ belong to a class of PPT state if the inequality (\ref{eqn:PPT_cond_simple}) holds. Hence proved.\\
It may be noted here that the quantity in the LHS of the inequality (\ref{eqn:PPT_cond_simple}) can be calculated if a few characteristic of the simple graph is known such as degree of each vertex and the total number of vertices in the graph.\\
Now, we take a few examples demonstrating the application of the above PPT condition given in \hyperref[thm:PPT_cond_alpha]{Theorem~(\ref*{thm:PPT_cond_alpha})}.\\ 
\textbf{Example-1:} Let us consider the complete graph $K_4$ given in Fig.(\ref{fig:K4}). 
The adjacency matrix $A_{K_4}$ and the corresponding density matrix $\rho_{\alpha}^{A_{K_4}}$ are as follows
\begin{eqnarray}
A_{K_4} =
\begin{bmatrix}
0 & 1 & 1 & 1 \\
1 & 0 & 1 & 1 \\
1 & 1 & 0 & 1 \\
1 & 1 & 1 & 0
\end{bmatrix}    
\end{eqnarray}
\begin{eqnarray}
\rho_\alpha^{A_{K_4}} = \frac{1}{12}
\begin{bmatrix}
3 & \beta & \beta & \beta \\[6pt]
\beta & 3 & \beta & \beta \\[6pt]
\beta & \beta & 3 & \beta \\[6pt]
\beta & \beta & \beta & 3
\end{bmatrix},\quad  \beta=\frac{1-\alpha}{\alpha}.
\end{eqnarray}
The $A_\alpha-$graph states described by the density matrix $\rho_\alpha^{A_{K_4}}$ is defined for $\frac{1}{4}\leq \alpha<1$ and it can be verified that  $\rho_\alpha^{A_{K_4}}$ represent a class of separable states when $\alpha \in [\frac{1}{4},1]$.\\
For $K_4$, we have $d_G=12$, $\sum_{v\in V(G)}d_v^2=36$ and $d_{1}=d_{2}=2$. Thus, the quantity in the LHS of the inequality (\ref{eqn:PPT_cond_simple}) can be calculated as 
\begin{align*}
    \frac{1}{1+\sqrt{\frac{d_G}{d_1 d_2-1}-\frac{1}{d_G}\sum_{v\in V(G)}d_v^2}}&=\frac{1}{2}
\end{align*}
Therefore, using the inequality given in (\ref{eqn:PPT_cond_simple}), we can say that $\rho_\alpha^{A_{K_4}}$ represent a class of separable states for $\frac{1}{2}\leq \alpha<1$ as the density matrix $\rho_{\alpha}^{(A_{G})}$ describe a two-qubit quantum state. It is worth noting that the condition given in (\ref{eqn:PPT_cond_simple}) is not a necessary condition for $\rho_{\alpha}^{(A_{G})}$ to be a class of PPT states, it is only a sufficient one. While this condition correctly identifies the class of states $\rho_{\alpha}^{(A_{K_4})}$ as a class of separable states for $\frac{1}{2}\leq \alpha<1$ but it does not detect all separable states belong to the family.\\
\textbf{Example-2:} Let us now consider the path graph $P_6$ given in Fig.5. 
\begin{figure}[h]
\centering
\begin{tikzpicture}
    \node[circle, draw] (1) at (0,2) {$v_{00}$};
    \node[circle, draw] (2) at (2,2) {$v_{01}$};
    \node[circle, draw] (3) at (4,2) {$v_{02}$};
    \node[circle, draw] (4) at (0,0) {$v_{10}$};
     \node[circle, draw] (5) at (2,0) {$v_{11}$};
     \node[circle, draw] (6) at (4,0) {$v_{12}$};

    \draw (1) -- (2) node[midway, above]{};
     \draw (2) -- (3) node[midway, left]{};
      \draw (3) -- (4) node[pos=0.25, above] {};
    \draw (4) -- (5) node[pos=0.75, above] {};
    \draw (5) -- (6) node[pos=0.75, above] {};
\end{tikzpicture}
\caption{Path graph $P_6$}
\label{fig:P6}
\end{figure}
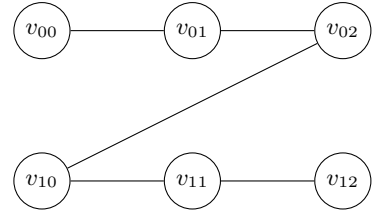
Its adjacency matrix $A_{P_6}$ and the corresponding $A_\alpha-$graph states described by the density matrix  $\rho_{\alpha}^{\left(A_{P_6}\right)}$ are given below
\begin{eqnarray}
A_{P_6} =
\begin{bmatrix}
0 & 1 & 0 & 0 & 0 & 0 \\
1 & 0 & 1 & 0 & 0 & 0 \\
0 & 1 & 0 & 1 & 0 & 0 \\
0 & 0 & 1 & 0 & 1 & 0 \\
0 & 0 & 0 & 1 & 0 & 1 \\
0 & 0 & 0 & 0 & 1 & 0
\end{bmatrix}
\end{eqnarray}
\begin{eqnarray}
\rho_\alpha^{(A_{P_6})} = \frac{1}{16}
\begin{bmatrix}
1 & \beta & 0 & 0 & 0 & 0 \\
\beta & 2 & \beta & 0 & 0 & 0 \\
0 & \beta & 2 & \beta & 0 & 0 \\
0 & 0 & \beta & 2 & \beta & 0 \\
0 & 0 & 0 & \beta & 2 & \beta \\
0 & 0 & 0 & 0 & \beta & 1
\end{bmatrix}, \quad \beta=\frac{1-\alpha}{\alpha}
\end{eqnarray}
The class of states $\rho_\alpha^{(A_{P_6})}$ is defined for $\frac{1}{2}\leq \alpha <1$.
From the condition (\ref{eqn:PPT_cond_simple}), we obtain that $\rho_\alpha^{(A_{P_6})}$ represent a class of separable states for $0.691 \leq \alpha <1$.
\begin{corollary}
    Let $G$ be a simple graph, $A_G$ be its adjacency matrix, and $\rho_{\alpha}^{A_G}$ be the $A_\alpha-$graph state corresponding to the graph $G$. If
    $\rho_{\alpha}^{A_G}$ is a class of negative partial transpose entangled states then
    \[\alpha_0 \leq \alpha < \Bigg[\frac{1}{1+\sqrt{\frac{d_G}{d_1 d_2-1}-\frac{1}{d_G}\sum_{v\in V(G)}d_v^2}}\Bigg]
    \]
    where $\alpha_{0}=\frac{\lambda_{min}(A_{G})}{\lambda_{min}(A_{G})-\delta}$. 
\end{corollary}
Therefore, we have established a condition on the mixing parameter $\alpha$ for which $A_{\alpha}-$graph states $\rho_{\alpha}^{A_G}$ form a class of PPT states. The condition depends only on the degree of each vertex and the total number of vertices in the graph $G$.
\section{Graph Theoretic Formulation of Moments-Based Entanglement Condition for $A_\alpha-$graph States} 
\label{Sec:Graph_Moments_Based_Entanglement}
Let us start this section with the detection of entanglement by using the criterion based on the moments of the partial transposition of the density matrix. Since partial transposition operation is a positive but not completely positive map, it is not physical and thus it may not be implemented in the experiment. But despite of the above difficulty in realizing the partial transposition operation in the experiment, the measurement of their moments is possible. This fact has been observed in many research. In one of the methods known as classical shadow formalism that allows for reliably estimating moments from randomized single-qubit measurements \cite{Huang2020}. Another way of measuring moments is using the cyclic shift operators on the multiple copies of of a state represented by an $m \times m$ density matrix $\rho$. It has also been shown that measuring partial moments is technically possible using $m$ copies of the state and controlled swap operations \cite{Ekert2002, Elben2018}. Recently, a method based on machine learning for measuring moments of any order has also been proposed \cite{Gray2018,Lin2023}. Therefore, since the estimation of the moments of the partial transposition are possible in the experiment so Elben et  al proposed a method for detecting bipartite entanglement in a many-body mixed state based on the estimation of the second and third moments of the partial transposition of the density matrix \cite{elben2020}. The above mentioned entanglement detection criterion is known as $p_3$-PPT criterion. The $p_3$-PPT condition states that any $m \otimes n$ dimensional bipartite PPT state described by the density operator $\rho_{PPT}$ lying in the composite system $XY$ satisfies the following inequality
\begin{eqnarray}
(p_2(\rho_{PPT}^{T_{Y}}))^2 - p_3(\rho_{PPT}^{T_{Y}}) \leq 0 \label{p3ppt1}
\end{eqnarray}
where $p_2(\rho_{PPT}^{T_{Y}})=Tr[(\rho_{PPT}^{T_{Y}})^{2}]$
and $p_3(\rho_{PPT}^{T_{Y}})=Tr[(\rho_{PPT}^{T_{Y}})^{3}]$.
\\
\subsection{$p_3$-PPT condition for $A_\alpha-$graph states}
In this section, our aim is to establish $p_3$-PPT condition in a graph theoretical way. By doing so, we can visualise the condition through graph and thus it may be provide us a simple way to extend the $p_3$-PPT condition for the multipartite system. We will use the properties of a graph to derive the $p_3$-PPT condition for $A_\alpha-$graph states.\\ 
To begin with, let us consider the $m\otimes n$ dimensional $A_\alpha-$ graph states corresponding to the graph $G$ and recall the second order moment of the partial transposition of $A_\alpha-$ graph states with respect to the subsystem $Y$ given in (\ref{eqn:p2}) as
\begin{eqnarray}
p_2((\rho_{\alpha}^{A_G})^{T_{Y}})= \frac{1}{(d_G)^2} \left[
    \sum_{v\in V(G)} (d_v)^2
    +\left(\frac{1-\alpha}{\alpha}\right)^2\|A_{G}\|_F^2\right]\nonumber\\
\label{p2}
\end{eqnarray}
Further, the third order moment of the partial transposition of $A_\alpha-$ graph states with respect to the subsystem $Y$ can be derived as (see Appendix)
\begin{widetext}
\begin{eqnarray}
 p_3((\rho_{\alpha}^{A_G})^{T_{Y}})= Tr[((\rho_{\alpha}^{A_G})^{T_{Y}})^{3}]=\frac{1}{(d_G)^3}\Bigg[&&\sum_{v\in V(G)} (d_v)^3+ 3\left(\frac{1-\alpha}{\alpha}\right)^2\,  \sum_{(v_{ik},v_{jl})\in E(G)}\left(d_{v_{il}}+d_{v_{jk}}\right)
(w_{v_{ik}v_{jl}})^2 +\nonumber\\&&6\left(\frac{1-\alpha}{\alpha}\right)^3 \sum_{\triangle \in \mathcal{T}(G^{T_B})} \prod_{(u,v) \in \triangle} w_{uv}\Bigg] 
\label{p3}
\end{eqnarray}
where $v_{ik}$ represents the $k^{th}$ vertex of $V_i$, $i\in\{A, B\}$, and $\mathcal{T}(G^{T_B})$ is the set of all triangles (undirected 3-cycles) in the partial transpose graph $G^{T_B}$, and $\triangle$ represents a triangle in the partial transpose graph $G^{T_B}$.
Now, if we assume that $\rho_{\alpha}^{A_G}$ belongs to a class of PPT states then $p_3$-PPT condition on the state $\rho_{\alpha}^{A_G}$ implies the following
\begin{eqnarray}
(p_2((\rho_{\alpha}^{A_G})^{T_{Y}}))^2\leq p_3((\rho_{\alpha}^{A_G})^{T_{Y}})   
\label{p3ppt}
\end{eqnarray}
Using (\ref{p2}) and (\ref{p3}) in the inequality (\ref{p3ppt}), we get
\begin{align*}
\left(
\sum_{v\in V(G)} (d_v)^2
    + \left(\frac{1-\alpha}{\alpha}\right)^2\|A_{G}\|_F^2\right)^2 
     \leq 
    d_G \Bigg[&\sum_{v\in V(G)} (d_v)^3+ 3\left(\frac{1-\alpha}{\alpha}\right)^2\, \sum_{(v_{ik},v_{jl})\in E(G)}\left(d_{v_{il}}+d_{v_{jk}}\right)
(w_{v_{ik}v_{jl}})^2\\
    &\quad + \left(\frac{1-\alpha}{\alpha}\right)^3 6 \sum_{\triangle \in \mathcal{T}(G^{T_B})} \prod_{(u,v) \in \triangle} w_{uv}\Bigg]
\end{align*}  
\end{widetext}
Therefore, the above result can be summarised by the theorem stated as follows:\\
\begin{theorem}
\label{thm:graphical_PPT_cond}
Let G be a weighted graph, and $A_G$ be its weighted adjacency matrix. Let $(V_A, V_B)$ be a fixed bipartition of the vertex set $V(G)$. If $\rho_{\alpha}^{A_{G}} \in PPT$ for some $\alpha \in I \subseteq (\alpha_{0},1]$, where $\alpha_{0}=\frac{\lambda_{min}(A_{G})}{\lambda_{min}(A_{G})-\delta}$, then the following inequality holds
\begin{widetext}
\begin{equation}
        \label{eqn:graphical_PPT_cond}
        \begin{split}
    \left(
    \sum_{v\in V(G)} (d_v)^2
    + \left(\frac{1-\alpha}{\alpha}\right)^2\|A\|_F^2\right)^2 
     \leq 
    d_G \Bigg[&\sum_{v\in V(G)} (d_v)^3+ 3\left(\frac{1-\alpha}{\alpha}\right)^2\, \sum_{(v_{ik},v_{jl})\in E(G)}\left(d_{v_{il}}+d_{v_{jk}}\right)
(w_{v_{ik}v_{jl}})^2\\
    &\quad + 6 \left(\frac{1-\alpha}{\alpha}\right)^3 \sum_{\triangle \in \mathcal{T}(G^{T_B})} \prod_{(u,v) \in \triangle} w_{uv}\Bigg]
     \end{split}
    \end{equation}    
\end{widetext}
\end{theorem}
where $v_{ik}$ represents the $k^{th}$ vertex of $V_i$, $i\in\{A,B\}$, and $\triangle_{G^{T_B}}$ is the number of triangles in the partial transpose graph $G^{T_B}$ of the graph $G$.
\begin{corollary}
    \label{cor:graph_ent_cond}
     Let G be a weighted graph, and $A_G$ be its weighted adjacency matrix. If equation (\ref{eqn:graphical_PPT_cond}) is violated by $\rho_{\alpha}^{A_{G}}$ for some $\alpha \in I \subseteq (\frac{\lambda_{min}(A_{G})}{\lambda_{min}(A_{G})-\delta},1]$, then $\rho_{\alpha}^{A_{G}}$ would represent a class of entangled states when the mixing parameter 
     $\alpha \in I$, 
\end{corollary}
\subsection{Examples-II}
We are now in a position to detect the PPT $A_{\alpha}-$ graph state using \hyperref[thm:graphical_PPT_cond]{Theorem~(\ref*{thm:graphical_PPT_cond})}. In $2\otimes 2$ system, PPT states are indeed a separable states and therefore, violation of \hyperref[thm:graphical_PPT_cond]{Theorem~(\ref*{thm:graphical_PPT_cond})} indicate the fact that the $A_{\alpha}-$ graph state under consideration is a class of entangled states. We will now verify the statements made with a few examples.
\begin{example}
    Let us consider the path graph $P_4$. The path graph $P_4$, its adjacency matrix $A_{P_4}$ and the corresponding $A_\alpha-$graph state $\rho_{\alpha}^{A_{P_4}}$ are as follows,
\begin{figure}[h]
\centering
\begin{tikzpicture}
    \node[circle, draw] (1) at (0,2) {$v_{00}$};
    \node[circle, draw] (2) at (2,2) {$v_{01}$};
    \node[circle, draw] (3) at (0,0) {$v_{10}$};
    \node[circle, draw] (4) at (2,0) {$v_{11}$};

    \draw (1) -- (2) node[midway, above] {};
    \draw (2) -- (3) node[midway, above]{} ;
    \draw (3) -- (4) node[midway, above] {};
\end{tikzpicture}
\caption{Path graph $P_4$}
\label{fig:P4}
\end{figure}
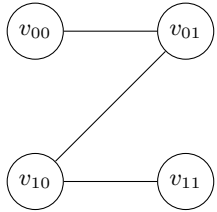
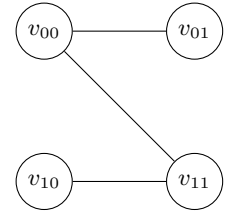
\begin{figure}[h]
\centering
\begin{tikzpicture}
    \node[circle, draw] (1) at (0,2) {$v_{00}$};
    \node[circle, draw] (2) at (2,2) {$v_{01}$};
    \node[circle, draw] (3) at (0,0) {$v_{10}$};
    \node[circle, draw] (4) at (2,0) {$v_{11}$};

    \draw (1) -- (2) node[midway, above] {};
    \draw (1) -- (4) node[midway, above] {};
    \draw (3) -- (4) node[midway, above] {};
\end{tikzpicture}
\caption{Partial transpose graph $P_4^{T_B}$ of the Path graph $P_4$}
\label{fig:P4TB}
\end{figure}
\[
A_{P_4} =
\begin{bmatrix}
0 & 1 & 0 & 0 \\
1 & 0 & 1 & 0 \\
0 & 1 & 0 & 1 \\
0 & 0 & 1 & 0
\end{bmatrix}
\]
\[
\rho_\alpha^{A_{P_4}} = \frac{1}{6}
\begin{bmatrix}
1 & \beta & 0 & 0 \\[6pt]
\beta & 2 & \beta & 0 \\[6pt]
0 & \beta & 2 & \beta \\[6pt]
0 & 0 & \beta & 1
\end{bmatrix},\quad  \beta=\frac{1-\alpha}{\alpha}.
\]
Here, $\rho_{\alpha}^{(A_{P_4})}$ is defined for $0.5\leq \alpha <1$.
The partial transpose graph $P_4^{T_B}$ corresponding to the graph $P_4$ is given by FIG. (\ref{fig:P4TB}).\\

For $P_4$, $d_G=6$, $\sum_{v\in V(G)} (d_v)^2=10$, and $\|A\|_F^2=6$.\\
Therefore, the left-hand side of inequality (\ref{eqn:graphical_PPT_cond}) is given by,
\begin{equation*}
    \left(
    \sum_{v\in V(G)} (d_v)^2
    + \left(\frac{1-\alpha}{\alpha}\right)^2\|A\|_F^2\right)^2 =\left(10+6 \left(\frac{1-\alpha}{\alpha}\right)^2\right)^2
\end{equation*}
The right-hand side of equation (\ref{eqn:graphical_PPT_cond}) is given by,
\begin{widetext}
\begin{align*}
    d_G \Bigg[\sum_{v\in V(G)} (d_v)^3+ 3\left(\frac{1-\alpha}{\alpha}\right)^2\, \sum_{(v_{ik},v_{jl})\in E(G)}\left(d_{v_{il}}+d_{v_{jk}}\right)
    &+ 6 \left(\frac{1-\alpha}{\alpha}\right)^3 \sum_{\triangle \in \mathcal{T}(G^{T_B})} \prod_{(u,v) \in \triangle} w_{uv}\Bigg]\\
    &= 6\Bigg[18+ 3\left(\frac{1-\alpha}{\alpha}\right)^2 8+ 0\Bigg]\\
    &= \Bigg[108+ 144\left(\frac{1-\alpha}{\alpha}\right)^2\Bigg]
\end{align*}
\end{widetext}
Now, inequality-(\ref{eqn:graphical_PPT_cond}) reduced to,
\begin{align}
    \left(10+6 \left(\frac{1-\alpha}{\alpha}\right)^2\right)^2&\leq \Bigg[108+ 144\left(\frac{1-\alpha}{\alpha}\right)^2\Bigg]
\end{align}
This inequality is not satisfied for $0<\alpha< 0.5117$. Thus, from corollary (\ref{cor:graph_ent_cond}), $\rho_{\alpha}^{(A_{P_4})}$ is a family of entangled states where $0.5 \leq\alpha < 0.5116$. This result aligns with the  Peres-Horodecki PPT criterion, which detects the class of states $\rho_{\alpha}^{(A_{P_4})}$ as entangled for $0.5 \leq\alpha < 0.5773$.
\end{example}
Thus, in this section, we obtained the inequality (\ref{eqn:graphical_PPT_cond}), which is the graph theoretical formulation of the moments-based PPT entanglement criterion given by Elben et al. \cite{elben2020}.

\begin{corollary}
\label{cor:graphical_PPT_cond_simple}
    Let G be a graph, and $A_G$ be its adjacency matrix. Let $(V_A, V_B)$ be a fixed bipartition of the vertex set $V(G)$. If $\rho_{\alpha}^{A_{G}} \in PPT$ for $\alpha \in I \subseteq (0,1]$ then
    
    \begin{equation}
        \label{eqn:graphical_PPT_cond-simple}
        \begin{split}
    &\left(
    \sum_{v\in V(G)} (d_v)^2
    + \left(\frac{1-\alpha}{\alpha}\right)^2 d_G\right)^2 
     \leq\\ 
   & d_G \Bigg[\sum_{v\in V(G)} (d_v)^3+ 3\left(\frac{1-\alpha}{\alpha}\right)^2\, \sum_{(v_{ik},v_{jl})\in E(G)}\left(d_{v_{il}}+d_{v_{jk}}\right)\\
    &\qquad + 6 \left(\frac{1-\alpha}{\alpha}\right)^3 \triangle_{G^{T_B}} \Bigg]
     \end{split}
    \end{equation}
\end{corollary}
where $v_{ik}$ represents the $k^{th}$ vertex of $V_i$, $i\in\{A,B\}$, and $\triangle_{G^{T_B}}$ is the number of triangles in the partial transpose graph $G^{T_B}$ of the graph $G$.

\textbf{Proof:} For simple graphs, $\|A\|_F^2=d_G$, $w_{uv}=1$ for all $(u,v) \in E(G)$, which implies the weight of each edge in $G^{T_B}$ would also be 1 i.e. $w'_{uv}=1$ for all $(u,v)\in G^{T_B}$.
Now, $\sum_{\triangle \in \mathcal{T}(G^{T_B})} \prod_{(u,v) \in \triangle} w_{uv}$ represent the sum, over all triangles in $G^{T_B}$, of the products of the edge weights within each triangle. The products of the edge weights within each triangle would be 1, as the weight of each edge is 1 in each triangle. Thus,
\begin{align*}
   \sum_{\triangle \in \mathcal{T}(G^{T_B})} \prod_{(u,v) \in \triangle} w_{uv}&=\sum_{\triangle \in \mathcal{T}(G^{T_B})} 1\\
   \sum_{\triangle \in \mathcal{T}(G^{T_B})} \prod_{(u,v) \in \triangle} w_{uv}&= \text{Number of triangles in $G^{T_B}$}\\
   &=\triangle_{G^{T_B}}
\end{align*}
Therefore, by \hyperref[thm:graphical_PPT_cond]{Theorem (\ref{thm:graphical_PPT_cond})}
    \begin{align*}
        \label{eqn:graphical_PPT_cond_simple}
        \begin{split}
   & \left(
    \sum_{v\in V(G)} (d_v)^2
    + \left(\frac{1-\alpha}{\alpha}\right)^2 d_G\right)^2 
     \leq \\
    &d_G \Bigg[\sum_{v\in V(G)} (d_v)^3
     + 3\left(\frac{1-\alpha}{\alpha}\right)^2\, \sum_{(v_{ik},v_{jl})\in E(G)}\left(d_{v_{il}}+d_{v_{jk}}\right)\\ + & 6 \left(\frac{1-\alpha}{\alpha}\right)^3 \triangle_{G^{T_B}} \Bigg]
     \end{split}
    \end{align*}
\begin{corollary}
\label{cor:graphical_nec_PPT_cond_simple_trianglefree}
    Let G be a simple, unweighted graph with $n$ vertices and at most $n-1$ edges and let the partial transpose graph $G^{T_B}$ be connected. Let $A_G$ be the adjacency matrix of $G$. If $\rho_{\alpha}^{A_{G}} \in PPT$ for $\alpha \in I \subseteq (0,1]$, then, for $\alpha \in I$,
\begin{align*}
    \begin{split}
    \left(
    \sum_{v\in V(G)} (d_v)^2
    + \left(\frac{1-\alpha}{\alpha}\right)^2 d_G\right)^2 
     \leq 
    d_G \Bigg[\sum_{v\in V(G)} (d_v)^3\\
    \quad+ 3\left(\frac{1-\alpha}{\alpha}\right)^2\,\sum_{(v_{ik},v_{jl})\in E(G)}\left(d_{v_{il}}+d_{v_{jk}}\right)\Bigg]
     \end{split}
\end{align*}
\end{corollary}
\textbf{Proof:} Since the graph $G$ has $n$ vertices and at most $n-1$ edges, its partial transpose $G^{T_B}$ also has $n$ vertices and at most $n-1$ edges. Since $G^{T_B}$ is connected, then $G^{T_B}$ has exactly $n-1$ edges, which means $G^{T_B}$ is a tree. Hence, it is triangle-free. Therefore, by corollary (\ref{cor:graphical_PPT_cond_simple}), the result follows immediately.
\section{Conclusion}
\label{Sec:Conclusion}
In this work, we have introduced $A_\alpha$-graph states, a new class of quantum states obtained directly from the adjacency and degree matrices of unweighted or weighted graphs. These states are density operators whose structure encodes the weighted connectivity of a graph via a tunable mixing parameter $\alpha$. We have shown that identifying the subinterval of $\alpha \in (0,1]$ ensuring positivity of $\rho_\alpha^{A_G}$ is a key step in establishing their physical validity.
Building on this construction, we obtained some promising results which enable us to generate a class of entangled states using a graph. The results presented here highlight a new connection between graph theory and moments-based quantum entanglement detection. The proposed formalism opens several directions for future research, including the classification of entanglement properties in $A_\alpha$-graph states for specific graph families, extensions to multipartite settings, and potential applications in noisy intermediate-scale quantum (NISQ) devices where graph-structured interactions naturally arise.
\section{Data Availability Statement}
 Data sharing is not applicable to this article as no datasets were generated or analysed during the current study.

\section{Appendix}
\label{Sec:Appendix}
\subsection{Proof of \hyperref[lem:Tr(DA_TB_sq]{Lemma~\ref{lem:Tr(DA_TB_sq}}}
In this subsection we provide the proof of \hyperref[lem:Tr(DA_TB_sq]{Lemma~\ref{lem:Tr(DA_TB_sq}}.\\
\textbf{Proof:} 
\begin{align*}
Tr\left(D(A^{T_B})^2\right)&=\sum_{v_{ik}\in V(G)}[d_{v_{ik}} [(A^{T_B})^2]_{v_{ik}v_{ik}}]     
\end{align*}
where $[(A^{T_B})^2]_{v_{ik}v_{ik}}$ represents the diagonal entry of $(A^{T_B})^2$ corresponding to the vertex $v_{ik}$. Thus,
\begin{align*}
Tr\left(D(A^{T_B})^2\right)&=\sum_{v_{ik}\in V(G)}[d_{v_{ik}} \sum_{v_{jl}\in V(G)}([(A^{T_B})]_{v_{jl}v_{ik}})^2]\\
&=\sum_{v_{ik}\in V(G)}[d_{v_{ik}} \sum_{v_{jl}\in V(G)}(w_{v_{jl}v_{ik}})^2]
\end{align*}
where, the inner sum represents the square of the weights of all the edges incident on the vertex $v_{ik}$ and the outer sum runs over all the vertices of $V(G)$. If we look this sum from the perspective of the edges in $G^{T_B}$, then each edge $(v_{il}, v_{jk})\in E(G^{T_B})$ contribute $d_{v_{il}}(w_{v_{il}v_{jk}})^2+d_{v_{jk}}(w_{v_{il}v_{jk}})^2$ to the above sum i.e. the trace of $D(A^{T_B})^2$. Therefore, 
\begin{align*}
Tr\left(D(A^{T_B})^2\right)&=\sum_{(v_{il},v_{jk})\in E(G^{T_B})}\Bigg[d_{v_{il}}(w_{v_{il}v_{jk}})^2\\
&\qquad +d_{v_{jk}}(w_{v_{il}v_{jk}})^2\Bigg]\\
&=\sum_{(v_{il},v_{jk})\in E(G^{T_B})}\Bigg[(d_{v_{il}}+d_{v_{jk}})(w_{v_{il}v_{jk}})^2 \Bigg]
\end{align*}
Now, since each edge $(v_{il},v_{jk})\in E(G^{T_B})$ with non-zero weight becomes the edge $(v_{ik},v_{jl})\in E(G)$, therefore,
\begin{align*}
Tr\left(D(A^{T_B})^2\right)&=\sum_{(v_{ik},v_{jl})\in E(G)}(d_{v_{il}}+d_{v_{jk}})(w_{v_{ik}v_{jl}})^2
\end{align*}
\subsection{Third order moment}
In this subsection we present the graph theretical expression of the third order moment of the partial transpose of $A_{\alpha}$-graph state $\rho$.
\begin{widetext}
\begin{align*}
p_3(\rho^{T_B}) 
&= \operatorname{Tr}\!\Bigg[ \left( \rho^{T_B} \right)^3 \Bigg]\\
&=\operatorname{Tr}\Bigg[
        \frac{1}{(d_G)^3}\left(D + \frac{1-\alpha}{\alpha} A^{T_B}\right)^3\Bigg]\\
&= \frac{1}{(d_G)^3}\operatorname{Tr} \Bigg[ D^3
+ \left(\frac{1-\alpha}{\alpha}\right)\!\left( D^2 A^{T_B} + D A^{T_B} D + A^{T_B} D^2 \right) 
+ \left(\frac{1-\alpha}{\alpha}\right)^2 \left( D (A^{T_B})^2 + A^{T_B} D A^{T_B} + (A^{T_B})^2 D \right) \\
&\quad+ \left(\frac{1-\alpha}{\alpha}\right)^3 (A^{T_B})^3 \Bigg]\\
&= \frac{1}{(d_G)^3}\Bigg[ \operatorname{Tr} (D^3)
+ \left(\frac{1-\alpha}{\alpha}\right)\!\big( \operatorname{Tr} (D^2 A^{T_B}) +\operatorname{Tr}  (D A^{T_B} D)  + \operatorname{Tr} (A^{T_B} D^2) \big)+ \left(\frac{1-\alpha}{\alpha}\right)^2 \big( \operatorname{Tr} (D (A^{T_B})^2)\\
&\quad + \operatorname{Tr} (A^{T_B} D A^{T_B})+ \operatorname{Tr} ((A^{T_B})^2 D) \big)+\left(\frac{1-\alpha}{\alpha}\right)^3 \operatorname{Tr} (A^{T_B})^3\Bigg] \\
&= \frac{1}{(d_G)^3}\Bigg[\sum_{v\in V(G)} (d_v)^3+\left(\frac{1-\alpha}{\alpha}\right)\,0 + 3\left(\frac{1-\alpha}{\alpha}\right)^2\, \operatorname{Tr} \left(D (A^{T_B})^2 \right)+\left(\frac{1-\alpha}{\alpha}\right)^3 \operatorname{Tr} (A^{T_B})^3\Bigg]\\
&= \frac{1}{(d_G)^3}\Bigg[\sum_{v\in V(G)} (d_v)^3+ 3\left(\frac{1-\alpha}{\alpha}\right)^2\, \operatorname{Tr} \left(D (A^{T_B})^2 \right)+\left(\frac{1-\alpha}{\alpha}\right)^3 \operatorname{Tr} (A^{T_B})^3\Bigg]
\end{align*}
Now, applying \hyperref[lem:Tr(DA_TB_sq]{Lemma~(\ref{lem:Tr(DA_TB_sq})} and \hyperref[lemma:Tr_ATB_cube]{Lemma~(\ref{lemma:Tr_ATB_cube})},
\begin{equation}
\label{eqn:p3}
p_3(\rho^{T_B})=\frac{1}{(d_G)^3}\Bigg[\sum_{v\in V(G)} (d_v)^3+ 3\left(\frac{1-\alpha}{\alpha}\right)^2\, \sum_{(v_{ik},v_{jl})\in E(G)}\left(d_{v_{il}}+d_{v_{jk}}\right)
(w_{v_{ik}v_{jl}})^2 +6\left(\frac{1-\alpha}{\alpha}\right)^3 \sum_{\triangle \in \mathcal{T}(G^{T_B})} \prod_{(u,v) \in \triangle} w_{uv}\Bigg]
\end{equation}
\end{widetext}
\bibliographystyle{apsrev4-2}  
\bibliography{bibliography3}

@article{Peres1996,
   author = {Asher Peres},
   doi = {10.1103/PhysRevLett.77.1413},
   issn = {10797114},
   issue = {8},
   journal = {Physical Review Letters},
   month = {8},
   pages = {1413},
   publisher = {American Physical Society},
   title = {Separability Criterion for Density Matrices},
   volume = {77},
   url = {https://journals.aps.org/prl/abstract/10.1103/PhysRevLett.77.1413},
   year = {1996}
}

@article{Horodecki1996,
   author = {Michał Horodecki and Paweł Horodecki and Ryszard Horodecki},
   doi = {10.1016/S0375-9601(96)00706-2},
   issn = {0375-9601},
   issue = {1-2},
   journal = {Physics Letters A},
   month = {11},
   pages = {1-8},
   publisher = {North-Holland},
   title = {Separability of mixed states: necessary and sufficient conditions},
   volume = {223},
   url = {https://www.sciencedirect.com/science/article/pii/S0375960196007062},
   year = {1996}
}

@article{Elben2020,
   author = {Andreas Elben and Richard Kueng and Hsin Yuan (robert) Huang and Rick Van Bijnen et al.},
   doi = {10.1103/PhysRevLett.125.200501},
   issn = {10797114},
   issue = {20},
   journal = {Physical Review Letters},
   month = {11},
   pages = {200501},
   pmid = {33258654},
   publisher = {American Physical Society},
   title = {Mixed-State Entanglement from Local Randomized Measurements},
   volume = {125},
   url = {https://journals.aps.org/prl/abstract/10.1103/PhysRevLett.125.200501},
   year = {2020}
}

@article{Horn1985,
   author = {Roger A. Horn and Charles R. Johnson},
   doi = {10.1017/CBO9780511810817},
   journal = {Matrix Analysis},
   month = {12},
   publisher = {Cambridge University Press},
   title = {Matrix Analysis},
   year = {1985}
}

@article{Huang2020,
   author = {Hsin Yuan Huang and Richard Kueng and John Preskill},
   doi = {10.1038/s41567-020-0932-7},
   issn = {1745-2481},
   issue = {10},
   journal = {Nature Physics 2020 16:10},
   month = {6},
   pages = {1050-1057},
   publisher = {Nature Publishing Group},
   title = {Predicting many properties of a quantum system from very few measurements},
   volume = {16},
   url = {https://www.nature.com/articles/s41567-020-0932-7},
   year = {2020}
}

@article{Ekert2002,
   author = {Artur K. Ekert and Carolina Moura Alves and Daniel K. Oi and Micha� Horodecki and Pawe� Horodecki and L. C. Kwek},
   doi = {10.1103/PhysRevLett.88.217901},
   issn = {00319007},
   issue = {21},
   journal = {Physical Review Letters},
   month = {5},
   pages = {217901},
   publisher = {American Physical Society},
   title = {Direct Estimations of Linear and Nonlinear Functionals of a Quantum State},
   volume = {88},
   url = {https://journals.aps.org/prl/abstract/10.1103/PhysRevLett.88.217901},
   year = {2002}
}

@article{Elben2018,
   author = {A. Elben and B. Vermersch and M. Dalmonte and J. I. Cirac and P. Zoller},
   doi = {10.1103/PhysRevLett.120.050406},
   issn = {10797114},
   issue = {5},
   journal = {Physical Review Letters},
   month = {2},
   pages = {050406},
   pmid = {29481179},
   publisher = {American Physical Society},
   title = {Rényi Entropies from Random Quenches in Atomic Hubbard and Spin Models},
   volume = {120},
   url = {https://journals.aps.org/prl/abstract/10.1103/PhysRevLett.120.050406},
   year = {2018}
}

@article{Raussendorf2003,
   author = {Robert Raussendorf and Daniel E. Browne and Hans J. Briegel},
   doi = {10.1103/PHYSREVA.68.022312},
   issn = {10941622},
   issue = {2},
   journal = {Physical Review A - Atomic, Molecular, and Optical Physics},
   pages = {32},
   title = {Measurement-based quantum computation on cluster states},
   volume = {68},
   year = {2003}
}

@article{Dr2003,
   author = {W. Dür and H. Aschauer and H. J. Briegel},
   doi = {10.1103/PhysRevLett.91.107903},
   issn = {10797114},
   issue = {10},
   journal = {Physical Review Letters},
   month = {9},
   pages = {107903},
   publisher = {American Physical Society},
   title = {Multiparticle Entanglement Purification for Graph States},
   volume = {91},
   url = {https://journals.aps.org/prl/abstract/10.1103/PhysRevLett.91.107903},
   year = {2003}
}

@article{Schlingemann2001,
   author = {D. Schlingemann and R. F. Werner},
   doi = {10.1103/PhysRevA.65.012308},
   issn = {10941622},
   issue = {1},
   journal = {Physical Review A},
   month = {12},
   pages = {012308},
   publisher = {American Physical Society},
   title = {Quantum error-correcting codes associated with graphs},
   volume = {65},
   url = {https://journals.aps.org/pra/abstract/10.1103/PhysRevA.65.012308},
   year = {2001}
}

@article{Braunstein2006_1,
   author = {Samuel L. Braunstein and Sibasish Ghosh and Simone Severini},
   doi = {10.1007/S00026-006-0289-3/METRICS},
   issn = {02180006},
   issue = {3},
   journal = {Annals of Combinatorics},
   month = {12},
   pages = {291-317},
   publisher = {Springer},
   title = {The Laplacian of a graph as a density Matrix: A basic combinatorial approach to separability of mixed states},
   volume = {10},
   url = {https://link.springer.com/article/10.1007/s00026-006-0289-3},
   year = {2006}
}

@article{Cabello2014,
   author = {Adán Cabello and Simone Severini and Andreas Winter},
   doi = {10.1103/PhysRevLett.112.040401},
   issn = {00319007},
   issue = {4},
   journal = {Physical Review Letters},
   month = {1},
   pages = {040401},
   publisher = {American Physical Society},
   title = {Graph-Theoretic Approach to Quantum Correlations},
   volume = {112},
   url = {https://journals.aps.org/prl/abstract/10.1103/PhysRevLett.112.040401},
   year = {2014}
}

@article{Ray2021,
   author = {Maharshi Ray and Naresh Goud Boddu and Kishor Bharti and Leong-Chuan Kwek and Adán Cabello},
   doi = {10.1088/1367-2630/ABCACD},
   issn = {1367-2630},
   issue = {3},
   journal = {New Journal of Physics},
   month = {3},
   pages = {033006},
   publisher = {IOP Publishing},
   title = {Graph-theoretic approach to dimension witnessing},
   volume = {23},
   url = {https://iopscience.iop.org/article/10.1088/1367-2630/abcacd https://iopscience.iop.org/article/10.1088/1367-2630/abcacd/meta},
   year = {2021}
}

@article{Lockhart2018,
   author = {Joshua Lockhart and Otfried Gühne and Simone Severini},
   doi = {10.1103/PhysRevA.97.062340},
   issn = {24699934},
   issue = {6},
   journal = {Physical Review A},
   month = {6},
   pages = {062340},
   publisher = {American Physical Society},
   title = {Entanglement properties of quantum grid states},
   volume = {97},
   url = {https://journals.aps.org/pra/abstract/10.1103/PhysRevA.97.062340},
   year = {2018}
}

@article{Dutta2016,
   author = {Supriyo Dutta and Bibhas Adhikari and Subhashish Banerjee and R. Srikanth},
   doi = {10.1103/PhysRevA.94.012306},
   issn = {24699934},
   issue = {1},
   journal = {Physical Review A},
   month = {7},
   pages = {012306},
   publisher = {American Physical Society},
   title = {Bipartite separability and nonlocal quantum operations on graphs},
   volume = {94},
   url = {https://journals.aps.org/pra/abstract/10.1103/PhysRevA.94.012306},
   year = {2016}
}

@article{KUMAR2026131195,
title = {Construction of PPT entangled state and its detection by using second-order moment of the partial transposition},
journal = {Physics Letters A},
volume = {567},
pages = {131195},
year = {2026},
issn = {0375-9601},
doi = {https://doi.org/10.1016/j.physleta.2025.131195},
url = {https://www.sciencedirect.com/science/article/pii/S0375960125009752},
author = {Rohit Kumar and Satyabrata Adhikari}
}

@article{Gray2018,
   author = {Johnnie Gray and Leonardo Banchi and Abolfazl Bayat and Sougato Bose},
   doi = {10.1103/PhysRevLett.121.150503},
   issn = {10797114},
   issue = {15},
   journal = {Physical Review Letters},
   month = {10},
   pages = {150503},
   pmid = {30362777},
   publisher = {American Physical Society},
   title = {Machine-Learning-Assisted Many-Body Entanglement Measurement},
   volume = {121},
   url = {https://journals.aps.org/prl/abstract/10.1103/PhysRevLett.121.150503},
   year = {2018}
}

@article{Lin2023,
   author = {Xiaodie Lin and Zhenyu Chen and Zhaohui Wei},
   doi = {10.1103/PhysRevA.107.062409},
   issn = {24699934},
   issue = {6},
   journal = {Physical Review A},
   month = {6},
   pages = {062409},
   publisher = {American Physical Society},
   title = {Quantifying quantum entanglement via a hybrid quantum-classical machine learning framework},
   volume = {107},
   url = {https://journals.aps.org/pra/abstract/10.1103/PhysRevA.107.062409},
   year = {2023}
}

@book{west2000introduction,
  author    = {Douglas B. West},
  title     = {Introduction to Graph Theory},
  edition   = {2nd},
  publisher = {Prentice Hall},
  year      = {2000},
  isbn      = {0-13-014400-2}
}

@book{chung1997spectral,
  author    = {Fan R. K. Chung},
  title     = {Spectral Graph Theory},
  series    = {CBMS Regional Conference Series in Mathematics},
  volume    = {92},
  publisher = {American Mathematical Society},
  year      = {1997},
  isbn      = {0-8218-0315-8}
}

@article{Kumar2022,
   author = {Rohit Kumar and Satyabrata Adhikari},
   doi = {10.1088/1402-4896/aca22c},
   issue = {12},
   journal = {Physica Scripta},
   month = {11},
   pages = {125101},
   publisher = {IOP Publishing},
   title = {Detection of d1 ⨂ d2 dimensional bipartite entangled state: a graph theoretical approach},
   volume = {97},
   url = {https://dx.doi.org/10.1088/1402-4896/aca22c},
   year = {2022}
}

@article{Adhikari2017,
   author = {Bibhas Adhikari and Subhashish Banerjee and Satyabrata Adhikari and Atul Kumar},
   doi = {10.1007/S11128-017-1530-1},
   issn = {1573-1332},
   issue = {3},
   journal = {Quantum Information Processing 2017 16:3},
   month = {2},
   pages = {79-},
   publisher = {Springer},
   title = {Laplacian matrices of weighted digraphs represented as quantum states},
   volume = {16},
   url = {https://link.springer.com/article/10.1007/s11128-017-1530-1},
   year = {2017}
}
\end{document}